\documentstyle{mn}
\newcommand{\ratmas}{M_{{\mathit S}}/M_{{\mathit BH}}}
\newcommand{\ratmasinv}{M_{{\mathit BH}}/M_{{\mathit S}}}

\newcommand{\bh}{{\mathit BH}}
\newcommand{\gr}{{\mathit GR}}
\newcommand{\inin}{{\mathrm in}}
\newcommand{\ouou}{{\mathrm out}}
\newcommand{\Jseq}{J(d_G)}
\newcommand{\Eseq}{E(d_G)}
\newcommand{\bJseq}{\bar{J}(\bar{d}_G)}

\newcommand{\dg}{d_G}
\newcommand{\rse}{r_{{\mathit sec}}}
\newcommand{\rdy}{r_{{\mathit dyn}}}
\newcommand{\rR}{r_R}

\newcommand{\VC}{{\mathit VC}}
\newcommand{\tX}{{\tilde X}}
\newcommand{\tY}{{\tilde Y}}
\newcommand{\tZ}{{\tilde Z}}
\begin{document}

\title[The black hole--gaseous star close binary]
{Newtonian models for black hole--gaseous star close binary systems}
\author[K. Ury\=u and Y. Eriguchi]
{K\=oji Ury\=u$^{1,2}$ and Yoshiharu Eriguchi$^3$\\
$^1$International Center for Theoretical Physics, Strada Costiera 11, 
34100 Trieste, Italy\\
$^2$SISSA, Via Beirut 2-4, 34013 Trieste, Italy\\
$^3$Department of Earth Science and Astronomy,
Graduate School of Arts and Sciences,
University of Tokyo, \\Komaba, Meguro, Tokyo 153-8902, Japan}

\maketitle

\begin{abstract}
Circularly orbiting black hole--gaseous star close binary systems are 
examined by using { \it numerically exact} stationary configurations 
in the framework of Newtonian gravity.  We have chosen a polytropic
star for the fluid component of the binary system and considered two ideal 
situations: 1) a synchronously rotating star and 2) an irrotationally 
rotating star. They correspond to a rotating star under the influence of 
viscosity and to that in the inviscid limit, respectively.  By analyzing the 
stationary sequences of binary systems with small separations, we can discuss 
the final stages of black hole--gaseous star close binary systems.  Our 
computational results show that the binary systems reach the Roche(--Riemann) 
limit states or the Roche lobe filling states without suffering from 
hydrodynamical instability due to tidal force for a large parameter range of 
the mass ratio and the polytropic index.  It is very likely that such stable 
Roche(--Riemann) limits or Roche lobe filling states survive even under the 
general relativistic effect.  Therefore, at the final stage of the evolution 
which is caused by the emission of gravitational waves, the Roche overflow 
will occur instead of merging of a black hole and a star 
\end{abstract}

\section{Introduction}

Close binary systems containing a stellar mass ($\sim M_\odot$) black 
hole (BH hereafter) and a neutron star (NS) or a white dwarf (WD) would emit 
gravitational waves (GW).  As a result, the two component stars approach 
each other 
almost adiabatically.  At the final stage of such an evolution, binary 
systems would become sources of various astrophysical phenomena.  They are 
one of the most promising sources of gravitational wave emission, which will 
be detected by the ground based laser interferometric gravitational wave 
detectors (LIGO/VIRGO/TAMA/GEO, see e.g. Abramovici et al. 1992 and 
Cutler et al. 1993).  
They are also expected to be one of the possible sources of $\gamma$--ray 
bursts (GRB) (e.g. Paczy\'nski 1991).  

Because of the GW emission, the orbit of a binary system becomes circular 
and the separation of two components decreases adiabatically.  For the final 
states of such BH--star systems, we can expect several different possibilities 
(see e.g. Kidder, Will \& Wiseman 1992, Lai, Rasio \& Shapiro 1993a, and 
references therein).  One is 
inspiraling on a dynamical time scale as a result of the tidal or the general 
relativistic (GR) effect.  
As is well known, there is an innermost stable circular orbit (ISCO) for 
a test particle around the BH field such as the 
Schwarzschild solution.  Even if the GR effect is neglected, Newtonian 
tidal field makes the orbital motion unstable as shown in numerical 
simulations for binary star systems.  
Furthermore, the dissipation as a result of the GW emission 
produces a radial component of the velocity.  Consequently the binary system 
would begin to merge by inspiraling dynamically.  In other words, the fluid 
star would fall on to the black hole on a short 
dynamical time scale.  The other possibility is the Roche overflow, i.e. 
the mass transfer from the fluid star to the black hole.  When the star could 
approach a state in which the Roche lobe is filled up without suffering 
from dynamical instability, a certain amount of matter of the star would 
overflow from its Roche lobe to the black hole or environment.  Even after 
this mass overflow, the star and the black hole may still remain in a stable 
binary system or at least some amount of matter may orbit around the BH longer 
than the dynamical time scale (see Kochanek 1992, Bildsten \& Cutler 1992, 
and references 
therein).  These two possibilities are also suggested by recent hydrodynamical 
simulations for synchronously rotating binary systems \cite{kl97}.  

Configurations of such binary systems can be approximated well by 
quasi-stationary models even just before the merging process or the 
overflow process starts, because the time scale of the GW emission is much 
longer than the orbital period (see e.g. Shapiro \& Teukolsky 1983).  
Thus we can 
discuss final stages of the evolution of binary systems by computing 
sequences of stationary configurations (see e.g. Lai, Rasio \& Shapiro 1993b 
(LRS1 
hereafter) and 1994a (LRS2)).  Stationary sequences representing 
evolutionary tracks can be constructed as follows.  It is reasonable to 
assume that the masses of components are conserved during the evolution 
and that the equation of state (EOS) and the viscosity of the fluid star 
are fixed.  If the time scale of change due to viscosity is much shorter than 
that of the evolution due to the GW emission, the spin of the NS is 
synchronized to the orbital motion.  We call a point source 
and a corotating star a Roche type binary configuration.  On the other 
hand, there is another possibility that the binary configuration can tend to 
settle into a state in which the fluid star rotates with zero vorticity in the 
inviscid limit (Kochanek 1992 ; Bildsten \& Cutler 1992). We call it an 
irrotational 
Roche--Riemann (IRR) binary configuration.  These two types of configurations 
correspond to extended models of ellipsoidal equilibria for binary systems 
studied by Roche and by Aizenman to more realistic models in which deformation 
and compressibility of fluid are fully taken into account (see e.g. 
Chandrasekhar 1969 and Aizenman 1968).  

By using such stationary configurations, the final stages of the evolution 
of binary systems can be discussed.  As the separation of two objects 
decreases, the Roche or the IRR type binary system finally reaches a critical 
distance.  The critical distances for these systems are discussed in LRS1 
by using the ellipsoidal approximation for the fluid component.  The final 
fates of the BH-NS systems depend on the type of the critical state which the 
binary system first approaches.  As was stated in their paper, there appear to 
be three types of critical distances in the framework of Newtonian gravity.  
These critical states can be found by analyzing stationary sequences 
by expressing physical quantities as a function of the separation of 
the centers of the mass of binary components, $d_G$.  The turning point of 
physical quantities such as $J(d_G)$ or $E(d_G)$ corresponds to the point 
where some instability sets in, where $J$ and $E$ are the total angular 
momentum and the total energy, respectively.  

For the Roche type binaries, the secular instability will set in at the 
separation for which functions $J(d_G)$ and $E(d_G)$ reach their turning 
points which correspond to minima of these functions.  At this point, the 
rotation of a gaseous star will change secularly owing to the 
viscous effect.  The point where the dynamical instability will set in is 
expected to be located at the smaller separation than that of secular 
instability.  The component stars are thought to coalesce as a result 
of this dynamical instability at this point.  According to LRS1 
and the well-organized hydrodynamical simulations of synchronously 
rotating NS-NS binary systems, differences of values of these two distances 
are rather small and hence the hydrodynamical instability sets in around 
the states at the turning points of the curves of the physical quantities 
(see e.g. Rasio \& Shapiro 1992, 1994 ; Shibata, Oohara \& Nakamura 1997 ; 
but see also New \& Tohline 1997).  
We denote the separation at the turning point of $J(d_G)$ or 
$E(d_G)$ as $\rse$.  For the IRR type binaries, the turning point corresponds 
to the dynamical instability limit and there is no secular instability limit 
because the system is not subjected to the influence of the viscosity.  We 
denote the radius at the turning point for the IRR sequences as $\rdy$.  
Another critical distance for the BH-NS binary sequence is the 
Roche(--Riemann) limit where the separation of two components is a minimum 
or the Roche lobe filling state where the matter fills up its Roche lobe.  
We denote the corresponding separation as $r_R$.  

According to the results of LRS1 in which the ellipsoidal approximation 
of the polytropic star has been used, the relation $r_R < \rdy$ is always 
satisfied for the IRR binary systems. In other words, the IRR binary systems
at the Roche--Riemann limit are always dynamically unstable.  For the Roche 
binary systems, the conditions $r_R < \rse$ or $r_R < \rdy$ are satisfied for 
almost all polytropic indices $N$ and almost all mass ratios $\ratmas$ in 
which realistic values for the BH-NS binary systems are included (see also 
LRS2 section 2.4).  

In this paper, we investigate stationary sequences of the Roche and 
the IRR binary systems for distances near the critical radii.  We 
obtain {\it numerically exact} stationary configurations of the binary 
systems.  By using these sequences, we will show that, contrary to the 
results obtained from the ellipsoidal approximation, there exist dynamically 
stable Roche and IRR configurations in a wide range of parameters, i.e.
$r_R$ appears earlier than $\rse$ or $\rdy$ as the separation decreases.  
This is because those configurations around the critical radii, in particular 
at $r_R$, cannot be expressed well by the ellipsoidal approximation even for 
fairly stiff equations of state because of a large deformation of the stellar 
envelope due to compressibility (Ury\=u \& Eriguchi 1998a, Paper I hereafter, 
1998b).  
If the Roche(--Riemann) limit appears for the larger separation than the 
other hydrodynamical instability limit, the final stage of the evolution of 
binary systems would be drastically different from the previously suggested 
scenario. In other words, the Roche lobe overflow or the mass transfer from 
the star to the black hole is likely to occur instead of unstable plunge of 
the star to the black hole.  We will compute stationary sequences for various 
polytropic EOS's and mass ratios and reveal the interrelation of critical 
radii on the sequences.  Note that whether such a mass transfer proceeds 
stably or not should be examined carefully because it could depend 
crucially on the realistic EOS of the NS and so on.  

Concerning the gravitational field, it is still very difficult to take into 
account the GR effect to such binary systems which contain BHs.  As a first 
step of this kind of research, we only treat Newtonian gravity both for the 
black hole and for the fluid star.  However, we will estimate the GR effect 
and suggest that there exists a possibility of mass overflow from the star 
component for realistic values of the parameters of binary systems. 

Such systems consisting of a point source and a compressible fluid star have 
not been discussed yet for fully deformed configurations from the standpoint 
of stationary structures of binary systems.  Therefore our present results 
will be useful when results in Newtonian or GR hydrodynamical simulations 
of BH--NS or WD systems will be obtained.  

\section{Solving method for stationary configurations of BH--NS binary systems}

The solving method of synchronously rotating binary systems has been 
established by several authors (Hachisu \& Eriguchi 1984a, 1984b ; 
Hachisu 1986).  
Recently, the method for irrotational binary systems has also been developed 
by the present authors (Paper I).  Since detailed computational methods are 
described in these papers, we will briefly summarize the assumptions, the
formulation of the problem and the computational method below.  

\subsection{Assumptions}

As mentioned in Introduction, we consider binary systems consisting of a point 
source and a compressible fluid star.  We assume the polytropic relation for 
the EOS of the fluid star: 
\begin{equation}
p\,=\, K \rho^{1+1/N} = K \Theta^{N+1} \, ,
\end{equation}
where $p$, $\rho$, $\Theta$, $N$ and $K$ are the pressure, the density, 
the Emden function which is proportional to the enthalpy, the polytropic 
index and a constant, respectively.  The Emden function is used 
in the actual numerical computations.  However note that the Emden function 
defined here is not normalized as usual ones treated in spherical polytropic 
stars.  From the present knowledge of the neutron matter, the realistic NS's 
are approximated by polytropes with index $N=0.5\sim 1$, (see LRS2 
and references therein) and some WD's or low mass main sequence stars are 
approximated by polytropes with $N \sim 1.5$.  

As for the gravitational force, we employ Newtonian gravity.  
The gravitational potential $\phi = \phi_s+\phi_{\bh}$ consists of 
a contribution from the fluid star itself, $\phi_s$, and that from the BH, 
$\phi_{\bh}$.  The former is expressed in the integral form 
of the Poisson equation as follows: 
\begin{equation} \label{intpot}
  \phi_s({\bf r}) \, = \, - \, G
  \int_V {\rho({\bf r}^{'}) \over \left|\, {\bf r}-{\bf r}^{'} \,\right|}
  d^3 {\bf r}^{'},
\end{equation}
where $G$ is the gravitational constant and the integration is performed 
over the stellar interior $V$. For the latter potential we assume as follows:
\begin{equation} \label{pot}
\phi_{\bh}({\bf r}) = -\,{G M_{\bh} \over R}\ ,
\end{equation}
where $M_{\bh}$ is the mass of the black hole and $R$ is the distance from 
the BH to each fluid element of the star.  

Because of dissipation due to the GW emission, we may assume that the 
binary orbit becomes circular before the coalescence.  This assumption 
breaks down when the hydrodynamical instability or the GR effect produces a
radial infall velocity.  The radius where the infall velocity cannot be
ignored is that of the ISCO.  On the other 
hand, since we neglect all GR effects, we can construct exact circularly 
orbiting binary configurations even when the star fills the Roche lobe.  
Using these configurations we can discuss the `purely' hydrodynamical 
stability of the system.  Furthermore, if the star is not so relativistic, we 
can assume that the evolution due to the GW emission is adiabatic so that the 
binary can be regarded in a quasi-stationary state just before the 
coalescence \cite{st83}.  We will discuss the GR effect in a later 
section.  

We can also assume the existence of rotating frames of reference whose angular 
velocity relative to the inertial frame coincides with the orbital angular 
velocity of the binary system $\Omega$.  In this frame, shapes of binary stars
are static.  We use this rotating frame in the actual computations.  

Concerning the spin state of the fluid component, we consider the following
two cases.  One is a synchronously rotating state (Roche type).  In other 
words, every fluid element of star is rotating with the same constant angular 
velocity $\Omega$ around the axis of orbital rotation of the binary system.  
The other is an irrotational state in which the vorticity vanishes 
everywhere in the fluid in the inertial frame of reference (IRR type).  

\subsection{Formulation of the problem}

We use the spherical coordinates $(r,\theta,\varphi)$ whose origin is 
located at the geometrical center of the star. The distance from the
rotational axis to the origin, $d_c$, is expressed as
\begin{equation} \label{gdis}
d_c \, = \, {R_{\inin} \, + \, R_{\ouou} \over 2},
\end{equation}
where $R_{\inin}$ and $R_{\ouou}$ are distances from the rotational axis 
to the inner and the outer edges of the star (i.e. the points of the star 
nearest to the rotational axis and furthest from it), respectively.  The star
is assumed to be equatorially symmetric. Furthermore we assume symmetry 
about the plane which is defined by the rotational axis and the geometrical
center of the star.  Since the formulation must be different for the Roche 
type configurations and for the irrotational models, we will briefly explain 
two corresponding formulations.

\subsubsection{Basic equations for the Roche type binary systems}

For the Roche type binary system, the equation of continuity becomes 
trivial and the equation of motion is integrated to the following
Bernoulli's equation:
\begin{equation} \label{gberno}
-\,{1 \over 2}\,\varpi^2\,\Omega^2\,+\,K\,(N+1)\,\Theta\,+\,\phi\,=\,C \, ,
\end{equation}
where $\varpi$ is the distance from the rotational axis to the fluid 
particle and defined by 
\begin{equation} \label{length}
\varpi\,=\,\left\{d_c^2\,+\,
r^2\,\sin^2\theta\,+\,2\,d_c\,r\,\sin\theta\cos\varphi\right\}^{1/2}\, ,
\end{equation}
and $C$ an integration constant.  The Emden function can be obtained from 
equation (\ref{gberno}).

The Green's function $1/|{\bf r}-{\bf r}^{'}|$ 
which appears in equation (\ref{intpot}) can be expanded as follows: 
\begin{equation} \label{legen}
{1 \over \left|\, {\bf r}-{\bf r}^{'} \,\right|}\,=\,
\sum_{n=0}^\infty f_n(r,r')\,P_n(\cos\gamma),
\end{equation}
where $f_n(r,r')$ is defined as,
\begin{eqnarray}
\!\!\!\!\!\!\!
f_n(r,r')  =
\left\{
\begin{array}{cc}
\lefteqn{{1 \over r}\left({r' \over r}\right)^n \ ,
\quad {\rm for} \quad r' \le r \ ,}  \\ \\
\lefteqn{{1 \over r'}\left({r \over r'}\right)^n \ ,
\quad {\rm for} \quad r \le r' \ .} 
\end{array}
\right.
\end{eqnarray}
Here $P_n(\cos\gamma)$ is the Legendre function and $\gamma$ is the angle
between two position vectors ${\bf r}$ and ${\bf r'}$.

\subsubsection{Basic equations for the IRR type binary systems}

For the IRR binary system, we will solve for {\it stationary} structures 
because the star has non-zero spin even in the rotating frame.  Since we 
assume that the vorticity of the fluid component vanishes in the inertial 
frame, we can introduce the velocity potential $\Phi({\bf r})$ in the 
inertial frame as follows:
\begin{equation}
{\bf v} = \nabla \, \Phi \, ,
\end{equation}
where ${\bf v}$ is the velocity vector in the inertial frame.
The Euler equation of the fluid can be integrated to the generalized
Bernoulli's equation as follows in the rotating frame:
\begin{equation} \label{bern1}
 - \,({\bf \Omega} \times {\bf r})
\cdot \nabla \, \Phi \, + \, {1 \over 2} 
\left|\, \nabla \, \Phi \,\right|^2 \, + \,
\int{dp \over \rho} \, + \, \phi \, = \, C \ ,
\end{equation}
where ${\bf \Omega}$ is the orbital angular velocity vector of the binary 
which is identical to the angular velocity vector of the rotating frame 
relative to the inertial frame.  Here we have used the stationarity of the 
configuration in the rotating frame.  We note that the steady velocity
${\bf u}$ in the rotating frame is related to ${\bf v}$ as follows:
\begin{equation}
{\bf u} = {\bf v}\,-\,{\bf \Omega}\times{\bf r}\, .
\end{equation}

Since we assume configurations are in stationary states, the equation of 
continuity is expressed by using the velocity potential
as follows also in the rotating frame:
\begin{equation} \label{conti}
\nabla^2 \Phi \, = \, N ({\bf \Omega} \times {\bf r} \, - \,
\nabla \, \Phi) \cdot {\nabla \, \Theta \over \Theta} \ .
\end{equation}

For the IRR binary systems, the Emden function and the velocity potential are 
basic variables which can be obtained from  equations (\ref{bern1}) 
and (\ref{conti}).  The boundary conditions for these two variables are as 
follows:
\begin{eqnarray} \label{bcon}
(\nabla\,\Phi \,-\,{\bf \Omega}\times{\bf r}) \cdot {\bf n} \, = \, 0,
\quad & {\rm on\ the\ stellar\ surface,} \\
\Theta \, = \, 0, \quad & {\rm on\ the\ stellar\ surface.}
\end{eqnarray}

\subsection{Numerical scheme}

These basic equations are transformed into the surface fitted coordinate 
system for the IRR binary system as described in Paper I.  For the Roche 
binary system, we use the surface fitted coordinate system for models with 
polytropic index $0\le N \le 1$. For polytropes with $1 \le N$, we use 
the ordinary spherical coordinate system.  Stationary configurations are
obtained by using the SCF method developed by Ostriker \& Mark (1968).  
The detailed 
numerical method for the Roche binary system can be found in 
Hachisu~(1986) and that for the IRR binary system in Paper I.  

Coordinates are discretized as $(r_i,\theta_j,\varphi_k)$ (
$0 \le i \le N_r\, ,\  0 \le j \le N_\theta \ {\rm and} \
0 \le k \le N_\varphi$), where $(N_r,N_\theta,N_\varphi)$
are the numbers of grid points. 
For the Roche type binary systems, we have chosen $(N_r,N_\theta,N_\varphi)
=(12,8,16)$ for models in the surface fitted coordinates and 
$(N_r,N_\theta,N_\varphi)=(32,16,32)$ for models in the ordinary spherical 
coordinates.  Although the number of grid points for the surface fitted 
coordinates is smaller than that of the ordinary spherical coordinates, the 
surface fitted coordinate system attains higher accuracy and greater 
robustness.  
We have computed $N=1$ models by using both coordinate systems and checked 
that two results agree very well.  For example, differences of critical radii 
were less that $0.5 \%$.  
For the IRR type binary systems, since the equations are more complicated 
than those for the Roche type binary systems, we use 
$(N_r,N_\theta,N_\varphi)=(16,12,24)$ to keep higher accuracy.  We also take 
the number of the terms of the Legendre expansion $0 \le n \le N_l$ in the 
equation (\ref{legen}) up to $N_l=10$ for the Roche case with the surface 
fitted coordinate, and for other models $N_l=14$.  In the HSCF scheme, 
configurations are solved iteratively by starting from a certain 
initial guess.  We can obtain converged solutions after $15\sim40$ iterations 
for the Roche cases and $40\sim90$ iterations for the IRR cases.  

\section{Computational results}
\begin{table*}
\begin{minipage}{150mm}
\caption{
List of computed models for the Roche binary systems.  Characters in 
parentheses denote how the critical radii along 
each sequence appear (see text).  
}
\label{model-roc.tab}
\begin{tabular}{cllllllllllll}
\hline
%$\ratmas$ & & & & & $N$ & & & & & &  \\
$\ratmas$ & \multicolumn{10}{c}{$N$}  \\
\\[-2truemm]
1.0 & 0.0 (H) & --- & --- & 0.5 (H) & 0.6 (Hc) & 0.7 (Hc) & 0.8 (R) & --- 
& 1.0 (R) & 1.5 (R) \\
0.5 & 0.0 (H) & --- & --- & 0.5 (H) & 0.6 (Hc) & 0.7 (M) & 0.8 (R) & --- 
& 1.0 (R) & 1.5 (R)  \\
0.2 & 0.0 (H) & 0.3 (H) & 0.4 (H) & 0.5 (M) & 0.6 (R) & 0.7 (R) & 0.8 (R) 
& 0.9 (R) & 1.0 (R) & 1.5 (R)  \\
0.1 & 0.0 (H) & 0.3 (H) & 0.4 (H) & 0.5 (M) & 0.6 (R) & 0.7 (R) & 0.8 (R) 
& 0.9 (R) & 1.0 (R) & 1.5 (R)  \\
\hline
\end{tabular}
\end{minipage}
\end{table*}
\begin{table*}
\begin{minipage}{100mm}
\caption{
Physical quantities at critical radii for the Roche type binary 
systems.  For normalizations of quantities, see equations (\ref{pqnorm}), 
(\ref{pqdis}), (\ref{rnorm}) and (\ref{idnorm}).  For the models with 
$N=0$ and $0.5$, the first row corresponds to quantities at $\rse$ and the 
second row corresponds to those at $r_R$ for each mass ratio.  For the 
models with $N=1$ and $1.5$, quantities at $r_R$ are shown.
}
\label{roc-crit.tab}
\begin{tabular}{cccccccccccccc}
\hline
$\ratmas$ & $\hat{d}$ & $\hat{d}_G$ & $\bar{\Omega}$ & $\bar{J}$ &
$\bar{E}$ & $\bar{R}$ \\
\\[-2truemm]
\multicolumn{7}{c}{$N = 0$} \\
\\[-2truemm]
1.0 & 1.945 & 2.418 & 3.138(-1) & 1.391 & -7.377(-1) & 9.989(-1) \\
    & 1.587 & 2.316 & 3.413(-1) & 1.412 & -7.315(-1) & 9.989(-1) \\
0.5 & 1.837 & 2.450 & 3.079(-1) & 2.353 & -8.511(-1) & 9.989(-1) \\
    & 1.595 & 2.407 & 3.194(-1) & 2.372 & -8.461(-1) & 9.989(-1) \\
0.2 & 1.761 & 2.486 & 2.997(-1) & 4.552 & -1.115 & 9.989(-1) \\
    & 1.596 & 2.468 & 3.046(-1) & 4.569 & -1.111 & 9.989(-1) \\
0.1 & 1.709 & 2.492 & 2.974(-1) & 7.334 & -1.460 & 9.989(-1) \\
    & 1.619 & 2.482 & 2.998(-1) & 7.343 & -1.457 & 9.989(-1) \\
\\[-2truemm]
\multicolumn{7}{c}{$N = 0.5$} \\
\\[-2truemm]
1.0 & 1.746 & 2.267 & 3.453(-1) & 1.334 & -7.065(-1) & 1.023 \\
    & 1.587 & 2.238 & 3.534(-1) & 1.336 & -7.058(-1) & 1.026 \\
0.5 & 1.699 & 2.328 & 3.311(-1) & 2.268 & -8.266(-1) & 1.025 \\
    & 1.629 & 2.320 & 3.333(-1) & 2.269 & -8.264(-1) & 1.026 \\
0.2 & 1.623 & 2.368 & 3.215(-1) & 4.408 & -1.105 & 1.028 \\
    & 1.596 & 2.367 & 3.216(-1) & 4.409 & -1.105 & 1.029 \\
0.1 & 1.574 & 2.372 & 3.194(-1) & 7.115 & -1.468 & 1.031 \\
    & 1.574 & 2.372 & 3.194(-1) & 7.115 & -1.468 & 1.031 \\
\\[-2truemm]
\multicolumn{7}{c}{$N = 1.0$} \\
\\[-2truemm]
1.0 & 1.667 & 2.207 & 3.562(-1) & 1.286 & -6.616(-1) & 1.030 \\
0.5 & 1.664 & 2.281 & 3.382(-1) & 2.203 & -7.860(-1) & 1.030 \\
0.2 & 1.706 & 2.327 & 3.272(-1) & 4.312 & -1.072 & 1.032 \\
0.1 & 1.664 & 2.327 & 3.269(-1) & 6.979 & -1.446 & 1.035 \\
\\[-2truemm]
\multicolumn{7}{c}{$N = 1.5$} \\
\\[-2truemm]
1.0 & 1.667 & 2.201 & 3.557(-1) & 1.255 & -5.963(-1) & 1.040 \\
0.5 & 1.733 & 2.278 & 3.373(-1) & 2.170 & -7.211(-1) & 1.039 \\
0.2 & 1.761 & 2.322 & 3.274(-1) & 4.272 & -1.009 & 1.040 \\
0.1 & 1.709 & 2.317 & 3.282(-1) & 6.926 & -1.386 & 1.044 \\
\hline
\end{tabular}
\end{minipage}
\end{table*}

The stationary solutions have been computed by fixing the mass ratio $\ratmas$ 
and parameters for the EOS, i.e. $N$ and $K$, where $M_S$ is the mass of the 
fluid star.  Varying the separation of two components $\tilde{d}$, we can 
compute a series of stationary solutions.  We will call it a stationary 
sequence.  As mentioned above, we have considered synchronously rotating 
models (Roche type binary systems) and irrotational models (IRR type binary 
systems).  The former can be applied to the evolution of binary systems 
composed of a viscous fluid star, and the latter to that of an inviscid star. 

The normalized separation is defined as
\begin{equation} \label{sepa}
\tilde{d}\,=\,{d_c\,+\,d_{\bh} \over R_0}\, ,  
\end{equation}
where $d_{\bh}$ is the distance from the rotational axis to the center 
of the black hole.  The quantity $R_0$ is defined as 
\begin{equation} \label{rad}
R_0\, = \, {R_{\ouou} \, - \, R_{\inin} \over 2}.  
\end{equation}
We note that the BH is orbiting according to the Keplerian law.  

We tabulate several dimensionless quantities for stationary sequences 
around the critical radii.  Dimensionless quantities in those tables
are defined as follows:
%
%\begin{eqnarray} \label{pqnorm}
\[
\bar{d}_G\,=\,{d_G \over R_N} \, ,\quad
\bar{\Omega}\,=\,{\Omega \over (\pi G\bar\rho_N)^{1/2}}\, ,
\]
\begin{equation} \label{pqnorm}
\bar{J}\,=\,{J \over (GM_S^3R_N)^{1/2}}\, ,\quad {\rm and} \quad
\bar{E}\,=\,{T\,+\,W\,+\,U \over GM_S^2/R_N}\, ,\nonumber
\end{equation}
%\end{eqnarray}
%
where $R_N$ is the radius of the spherical polytrope with the same mass $M_S$ 
and the same polytropic index $N$.  The quantity $\bar{\rho}_N$ is defined as 
$\bar\rho_N\,=\,M/(4\pi R^3_N/3)$.  These normalizations are the same 
as those adopted in LRS's papers.  The quantities $T$, $W$ and $U$ are 
the total kinematic energy, the total potential energy and the total 
thermal energy defined as usual (see e.g. Tassoul 1978).
The separation between the mass centers of two components $d_G$ is
calculated as follows:
\begin{equation} \label{pqdis}
d_G \,=\,  {\int_V \,x\,\rho \,d^3 {\bf r}^{'} \over M_S}
 + d_{\bh} \ ,
\end{equation}
where $x$ is defined as $x=d_c + r\sin\theta \cos\varphi$.  
We also tabulate the quantity $T/\left| W \right|$.  (Note that $T$ 
includes both the kinetic energy of orbital motion and that of spin, 
and hence the definition is different from that of LRS1.)   The virial 
constant $\VC$ is also tabulated, which is normalized by the total 
gravitational potential as,  
\begin{equation} \label{virial}
\VC\,=\,{\left|\,2\,T\,+\,W\,+\,3\,U/N \,\right| \over 
\left|\, W \,\right|},
\end{equation}
where the term $U/N$ vanishes for models with $N=0$.
We also show values of the averaged radius $\bar{R}$ defined as follows:
\begin{eqnarray}\label{rnorm}
\bar{R} &=& \left({3\,V\over 4\,\pi\,R^3_N}\right)^{1\over 3} \nonumber\\
&=&
\left[{1\over4\,\pi\,R^3_N}\int_0^{\pi}\sin\theta\,d\theta \int_0^{2\pi}
d\varphi\,R^3(\theta,\varphi)\right]^{1\over 3}\ .
\end{eqnarray}
Generally, this quantity increases as the separation decreases.  It
implies that the volume of the gaseous star always increases and,
accordingly, that the central density decreases as the two components 
approach during the evolution.
In some tables, we also show quantities $\hat{d}$ and $\hat{d}_G$ which 
are defined as follows:
\[
\hat{d}\,=\,{d \over R_N (1 + \ratmasinv)^{1/3}}
\]
\begin{equation}\label{idnorm}
\hat{d}_G\,=\,{d_G \over R_N (1 + \ratmasinv)^{1/3}}.  
\end{equation}
$d$ behaves as $d \propto (\ratmas)^{-1/3}$ for $\ratmas \rightarrow 0$ limit 
since the binary system tends to Keplerian, which is the reason of this 
normalization (see LRS1).  
Finally, for later discussions, we write down the quadrupole formula 
for the gravitational radiation emission here.  In the rotating frame of 
reference, the quadrupole moment tensors are defined as
\begin{equation}
I_{ij}^{(rot)}\,=\,\rho\int_V(x_i x_j\,-\,\delta_{ij}{\mid {\bf x} \mid^2
\over 3})dV,
\end{equation}
where $i,j=1,2,3$ correspond to the $x,y,z$-component of the Cartesian
coordinates, respectively.  By taking the symmetry of the system into
account, the quadrupole formula for the angular momentum dissipation 
rate is written as follows:

\begin{equation} \label{quad}
{dJ \over dt}\,=\,-{32\,G \over 5\,c^5}\,\Omega^5\,(I_{11}^{(rot)}\,-\,
I_{22}^{(rot)})^2 \, ,
\end{equation}
where $c$ is the speed of light.  Derivation of this formula is shown
in the book of Misner, Thorne \& Wheeler~(1970) (see also Eriguchi,
Futamase \& Hachisu 1990).  
This angular momentum dissipation rate is related to 
the energy dissipation rate as follows:
\begin{equation}
{dE \over dt}\,=\,\Omega{d J \over dt}
\end{equation}
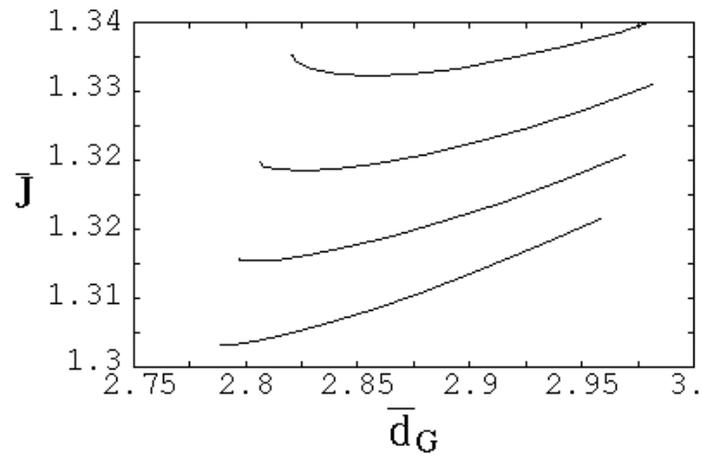
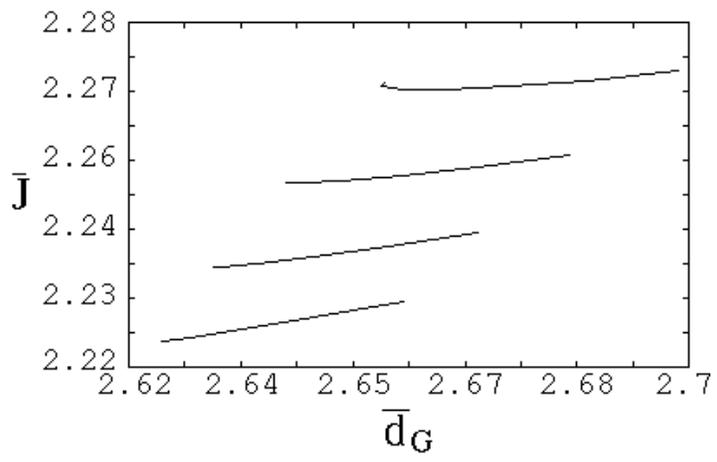
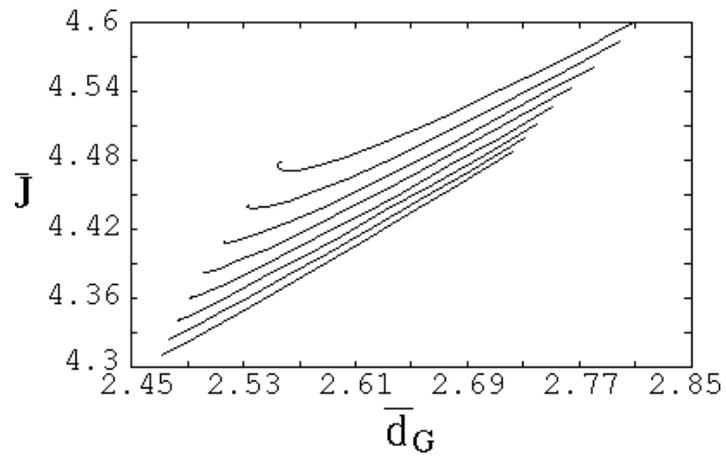
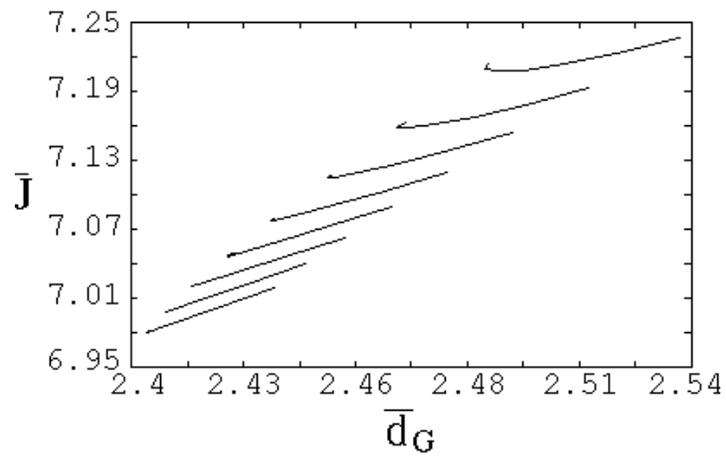
\begin{figure}
\vspace{20cm}
\caption{
Function $\bJseq$ of the Roche sequences for several mass ratios $\ratmas$
and polytropic indices $N$.
(a) $\ratmas = 1.0$ and $N=0.5$--$0.8$.
(b) $\ratmas = 0.5$ and $N=0.5$--$0.8$.
(c) $\ratmas = 0.2$ and $N=0.3$--$1.0$.
(d) $\ratmas = 0.1$ and $N=0.3$--$1.0$.
Each curve corresponds to a different polytropic index $N$.
The difference of $N$ between two adjacent curves is $0.1$.
The upper curve corresponds to the smaller value of $N$.}
\label{roc-d-J.gra}
\end{figure}
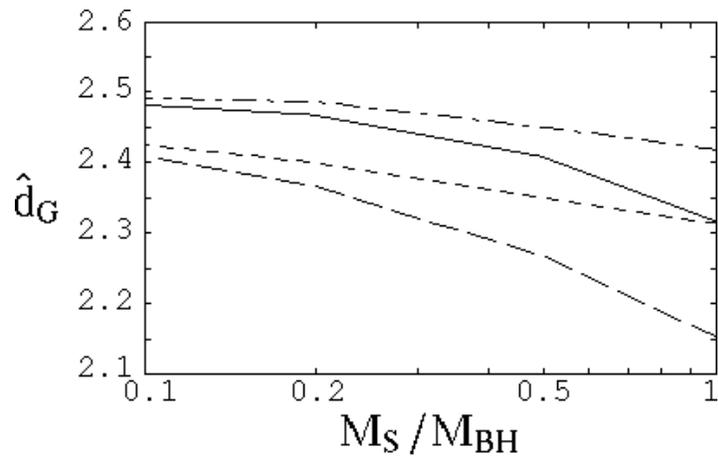
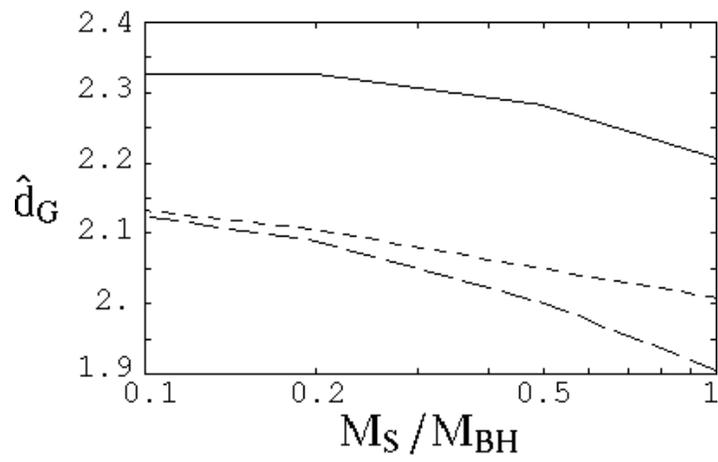
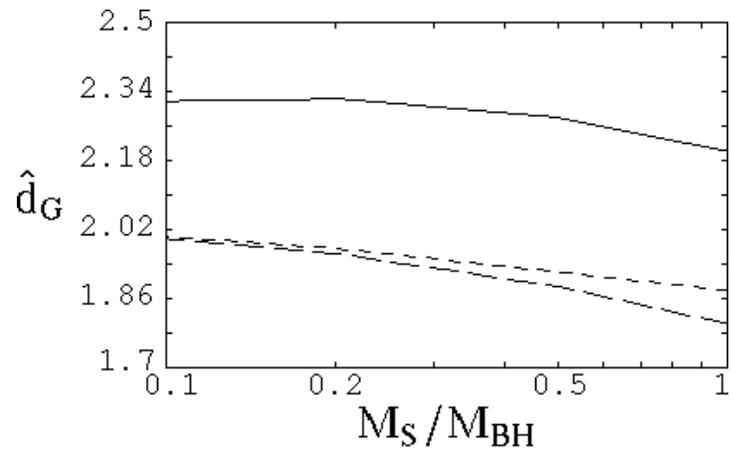
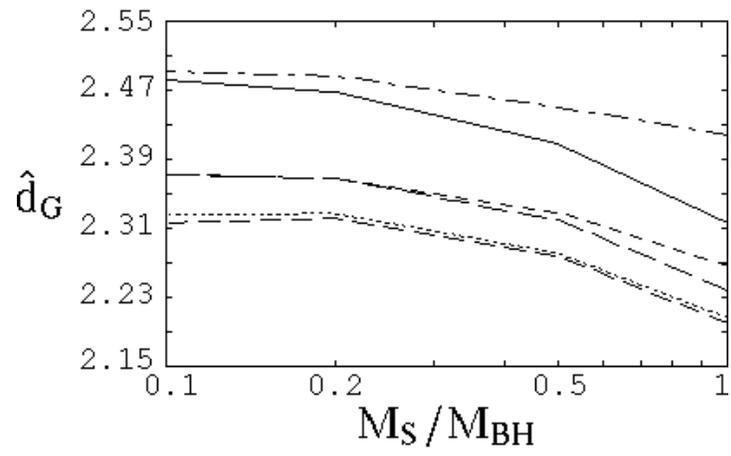
\begin{figure}
\vspace{19.5cm}
\caption{
Plots of critical radii $\rse$ and $\rR$ for the Roche sequences as a 
function of the mass ratio $\ratmas$.  In panels (a)--(c), present results 
and those of LRS1 are shown for the polytropic indices $N=0$, $1.0$ and $1.5$, 
respectively.  Each curve corresponds to a different critical radius as 
follows: present result for $\rse$ (dash dotted line), present result for 
$r_R$ (solid line), result of LRS1 for $\rse$ (short dashed line) and 
result of LRS1 for $r_R$ (long dashed line).  In the panel (d), present 
results for several polytropic indices $N = 0, 0.5, 1.0$ and $1.5$ are drawn. 
Dash dotted line: $\rse$ for $N=0$. 
Solid line: $r_R$ for $N=0$.
Short dashed line: $\rse$ for $N=0.5$.
Long dashed line: $r_R$ for $N=0.5$. 
Dotted line: $r_R$ for $N=1$.
Dashed line: $r_R$ for $N=1.5$. }
\label{roc-mass-radi.gra}
\end{figure}
\subsection{Results for the Roche type binary systems}

\subsubsection{Critical radii for the Roche type binary configurations}

As mentioned in Introduction, we can discuss the critical radii of 
binary systems by using the stationary sequences.  One can find a 
comprehensive 
discussion on these critical radii in LRS's papers in which the ellipsoidal 
approximation is used (see for example Lai, Rasio \& Shapiro 1993a, 1994b). 

According to LRS, three different types of critical radii can be identified 
by examining the behavior of $\Jseq$ or $\Eseq$ for the Roche type sequences.  
The turning point of the $\Jseq$ or $\Eseq$ curve corresponds to the point 
where the secular instability sets in.  We denote this separation as $\rse$.  
Since the assumption of the synchronous rotation implies that the viscosity of 
neutron matter is strong enough, viscosity can change the flow field of the 
fluid star and the star secularly evolves to a configuration with a lower 
energy even when the system is in a configuration at the turning point of 
the energy curve.  However, we also note that the evolutionary time scale due 
to the GW emission is almost comparable to the dynamical time scale at the 
configuration corresponding to the turning point of the energy curve.  

There is another critical radius on the sequence where the dynamical 
instability sets in, i.e. $\dg=\rdy$.  To determine $\rdy$, it is necessary 
to compute non-synchronously rotating configurations (see LRS1).  
However, it is difficult to construct such non-synchronously rotating 
configurations with generic flows (see for e.g. Ury\=u \& Eriguchi 1996, 
and references 
therein).  According to the ellipsoidal approximation of LRS and the dynamical 
simulations for the {\it NS-NS} binary systems (Rasio \& Shapiro 1992, 1994 ; 
Shibata, Oohara \& Nakamura 1997 ; Oohara, Nakamura \& Shibata 1997 ; 
Ruffert, Rampp \& Janka 1997 ; New \& Tohline 1997), $\rdy$ is smaller than 
that of the secular 
instability limit, $\rdy < \rse$, and these two radii are very close.  
For these kinds of {\it NS--NS} binary systems, they showed that 
the instability of orbital motion begins around these radii.  Since we can 
expect that the situation is similar for binary systems of a point source and 
a fluid star, we only discuss $\rse$ hereafter and regard it as the radius 
where the instability of orbital motion sets in.  We also refer to this 
instability, associated with inspiraling motions, as the hydrodynamical 
instability in this paper.  The origin of this instability is purely 
a Newtonian tidal effect for the fluid component.  

The other critical radius is the Roche limit $r_R$ which corresponds to 
the smallest separation of the stars along the sequences or the Roche lobe
filling states.  To determine this limit, we need to compute sequences by 
changing the parameter $\tilde{d}$ to smaller values until the inner edge 
of the fluid star forms a cusp.  The configuration with such a cusp can be 
identified as the Roche lobe filling state.  The smallest value of 
$d_G$ is the Roche limit $r_R=d_G$ of the sequence.  
It should be noted that it is possible for $r_R$ to appear at a larger 
separation than $\rdy$ or $\rse$ or both and hence the critical radius $\rdy$ 
or $\rse$ disappears.  In this case, by considering equation (\ref{quad}), 
the evolutionary time scale due to the GW emission does not become zero.  
Therefore the GW emission would not drive a significant radial velocity.  
If the GR effect is also negligible, the infall velocity would be a moderate 
value even at $r_R$.  Consequently, tidal stripping of the fluid component 
star or tidal disruption without the instability of orbital motion could be 
expected to occur.  

\subsubsection{Stationary sequences of the Roche type binary systems}
\begin{table*}
\begin{minipage}{130mm}
\caption{
List of computed models for the IRR binary systems.  Characters in 
parentheses denote how the critical radii along 
each sequence appear (see text).  
}
\label{model-irr.tab}
%\begin{tabular}{cccccccccccccc}
\begin{tabular}{cllllllllllll}
\hline
%$\ratmas$ & & & & $N$ & & & &   \\
$\ratmas$ & \multicolumn{7}{c}{$N$} \\
\\[-2truemm]
1.0 & 0.0 (H) & --- & --- & --- & 0.5 (Hc) & 0.6 (M) & 0.7 (R) & 1.0 (R) & \\
0.5 & 0.0 (H) & --- & --- & --- & 0.5 (Hc) & 0.6 (R)& 0.7 (R) & 1.0 (R) & \\
0.2 & 0.0 (H) & --- & 0.3 (Hc) & 0.4 (M) & 0.5 (R) & --- & --- & 1.0 (R) & \\
0.1 & 0.0 (H) & 0.2(Hc) & 0.3 (M) & 0.4 (R) & 0.5 (R) & --- & --- & 1.0 (R) & \\
\hline
\end{tabular}
\end{minipage}
\end{table*}
\begin{table*}
\begin{minipage}{100mm}
\caption{
Same as Table 2 but for IRR type binary systems.  For models with $N=0$ and 
$\ratmas=1.0$ and $0.5$ of $0.5$, the first row corresponds to quantities at 
$\rse$ and the second row corresponds to those at $r_R$ for each mass ratio.  
For models with $N=1$ and $1.5$ and $\ratmas=0.2$ and $0.1$ of $0.5$, 
quantities at $r_R$ are shown.  
}
\label{irr-crit.tab}
\begin{tabular}{cccccccccccccc}
\hline
$\ratmas$ & $\hat{d}$ & $\hat{d}_G$ & $\bar{\Omega}$ & $\bar{J}$ &
$\bar{E}$ & $\bar{R}$ \\
\\[-2truemm]
\multicolumn{7}{c}{$N = 0$} \\
\\[-2truemm]
1.0 & 1.905 & 2.415 & 3.152(-1) & 1.271 & -7.567(-1) & 9.994(-1) \\
    & 1.627 & 2.362 & 3.301(-1) & 1.287 & -7.506(-1) & 9.994(-1) \\
0.5 & 1.837 & 2.483 & 3.018(-1) & 2.244 & -8.678(-1) & 9.994(-1) \\
    & 1.664 & 2.450 & 3.101(-1) & 2.259 & -8.619(-1) & 9.994(-1) \\
0.2 & 1.761 & 2.518 & 2.944(-1) & 4.460 & -1.126 & 9.994(-1) \\
    & 1.651 & 2.509 & 2.969(-1) & 4.470 & -1.123 & 9.994(-1) \\
0.1 & 1.709 & 2.527 & 2.916(-1) & 7.258 & -1.466 & 9.994(-1) \\
    & 1.619 & 2.523 & 2.929(-1) & 7.268 & -1.463 & 9.994(-1) \\
\\[-2truemm]
\multicolumn{7}{c}{$N = 0.5$} \\
\\[-2truemm]
1.0 & 1.667 & 2.256 & 3.479(-1) & 1.223 & -7.240(-1) & 1.010 \\
    & 1.627 & 2.255 & 3.484(-1) & 1.223 & -7.239(-1) & 1.010 \\
0.5 & 1.699 & 2.336 & 3.291(-1) & 2.162 & -8.426(-1) & 1.011 \\
    & 1.664 & 2.335 & 3.295(-1) & 2.162 & -8.426(-1) & 1.012 \\
0.2 & 1.651 & 2.383 & 3.184(-1) & 4.302 & -1.121 & 1.014 \\
0.1 & 1.641 & 2.390 & 3.160(-1) & 7.009 & -1.484 & 1.015 \\
\\[-2truemm]
\multicolumn{7}{c}{$N = 1.0$} \\
\\[-2truemm]
1.0 & 1.706 & 2.205 & 3.560(-1) & 1.189 & -6.780(-1) & 1.005 \\
0.5 & 1.733 & 2.282 & 3.378(-1) & 2.109 & -8.012(-1) & 1.008 \\
0.2 & 1.761 & 2.328 & 3.273(-1) & 4.214 & -1.088 & 1.010 \\
0.1 & 1.754 & 2.337 & 3.250(-1) & 6.887 & -1.460 & 1.012 \\
\hline
\end{tabular}
\end{minipage}
\end{table*}

In Table \ref{model-roc.tab}, we tabulate model parameters chosen for the 
Roche type binary systems.  They are the mass ratio $\ratmas$ and the 
polytropic index $N$.  We have computed models with $0.1 \la \ratmas \la 1$ 
and $0 \la N \la 1.5$.  In Figure \ref{roc-d-J.gra}(a)--(d), we show $\bJseq$ 
for sequences with $\ratmas = 1$, $0.5$, $0.2$ and $0.1$, respectively, with 
several polytropic indices $N$ whose values are $0 < N < 1$.  Our results show 
important differences from those of LRS1.  In their results the secular 
instability limit $\rse$ always appears for any mass ratio and any polytropic 
index and always satisfies the relation $\rse > \rR$.  It can be seen that the 
turning point of the $\bJseq$ curve (i.e. $\rse$) disappears for $N \sim 0.8$, 
$0.7$, $0.5$ and $0.5$ with the mass ratio $\ratmas = 1$, $0.5$, $0.2$ and 
$0.1$, respectively.  

In Table \ref{model-roc.tab}, we roughly classify the sequences
by considering the appearance of critical radii along the curves of
$\Jseq$.  Three capital letters H, R and M are used as follows: 
\begin{enumerate}
\item[(i)] H denotes that the sequence has a turning point.  
\item[(ii)] R denotes that the sequence does not have a turning point but 
directly reaches a state with the smallest separation.  
\item[(iii)] M denotes that the sequence has a turning point which 
almost coincides with a state with the smallest separation.  
\end{enumerate}

As mentioned before, we have computed sequences by varying $\tilde{d}$ 
until the cusp is formed at the inner edge of the fluid star.  
For the H type sequences, some models at $r_R$ have cusps and others do not
have cusps.  A character c will be used for the models with cusps at $r_R$.  
For the R type sequences, the separations of the terminal configurations 
which fill up the Roche lobe should be the smallest.  However, as can be 
seen in Figure \ref{roc-d-J.gra}(a)--(d), the total angular momentum of the 
configuration with the cusp, i.e. the left end point of the curves $\Jseq$, 
slightly shows an irregular behaviour on the sequence in some cases.  
This may be because it is relatively difficult to maintain the same accuracy 
for those `singular' configurations with cusps as well as for the others.  
Since such irregular behaviour is very small, we can determine whether 
the sequence has a turning point or not from these figures.  
For example, the difference between $d_G$ of a configuration with a cusp 
and $r_R$ is less than $0.2\%$ for every sequence.  

According to the discussion in the previous section, the hydrodynamical 
instability is expected to occur for all models of (i), and the mass overflow 
is expected to occur prior to the orbital instability for models of (ii).  
Since the realistic NS is thought to be approximated by the polytropic EOS 
with $N \sim 0.5-1.0$ (see e.g. Shapiro \& Teukolsky 1983) and the WD or 
the low mass main 
sequence stars with $N \sim 1.5$, most of realistic situations of the BH--star 
binary systems correspond to the case (ii).  

We tabulate several physical quantities at critical radii, $\rse$ 
and $r_R$ in Table \ref{roc-crit.tab}.  
In Figure \ref{roc-mass-radi.gra}, we plot these critical radii 
against the mass ratio.  In panels (a)--(c), we compare our results 
with those of the ellipsoidal approximation tabulated in LRS1 for $N = 0$, 
$1.0$ and $1.5$.  For $N=0$, the differences of the values of 
critical radii are more or less $5 \%$ and stationary sequences 
behave qualitatively similarly for both results.   They can be classified
to the H type (see Ury\=u \& Eriguchi 1998c).  On the other hand, for 
larger $N$ 
the quantitative difference becomes larger.  It becomes roughly 
$10 \%$ for polytropes with $N=1$ and $15 \%$ for models with $N=1.5$.  
Moreover, the turning points disappear on the stationary sequences.
In Figure \ref{roc-mass-radi.gra}(d), we plot our results for polytropes
with $N = 0$, $0.5$, $1.0$ and $1.5$. From this figure, it can be
clearly shown that the secular instability limit vanishes for larger 
polytropic indices and for larger mass ratios as mentioned before.  
\subsection{Results for the IRR type binary systems}
\subsubsection{Critical radii for the IRR type binary systems}

The stationary sequence of the IRR type binary system can be regarded as 
that composed of an inviscid fluid star.  The circulation of each 
fluid element is conserved under the dissipation due to the GW emission 
\cite{mi74}.  Therefore, there is no secular instability limit on 
the IRR sequence but there are two different types of critical radii on it.  
The radius at the turning point of the $\Jseq$ or $\Eseq$ curve is denoted 
as $\rdy$.  The other critical radius is the Roche-Riemann limit $r_R$.  The 
determination of the Roche-Riemann limit is the same as the case of the Roche 
binary systems.  As discussed in LRS1 the dynamical instability will set in 
around $\rdy$.  On the other hand, if the $r_R$ is larger than $\rdy$, i.e. 
$\rdy$ disappears on the IRR sequence, the dynamical instability would not 
occur for binary systems during the evolution but the mass overflow 
would be expected to begin at $\rR$.  

\subsubsection{Stationary sequences of the IRR type binary systems}

In Table \ref{model-irr.tab}, we tabulate model parameters chosen for the IRR 
type binary systems.  The IRR configurations correspond to the inviscid limit 
of fluid stars which can be applied for the BH--NS systems, because the 
viscosity is strong enough to synchronize the rotation only for WD or lower 
main sequence stars.  Therefore we have computed polytropes with indices 
$0 \le N \le 1$, which are appropriate for the neutron matter.

In Figure \ref{irr-d-J.gra}(a)--(d), we show the quantity $\bJseq$ for 
sequences with $\ratmas = 1$, $0.5$, $0.2$ and $0.1$, respectively.
The dynamical instability limit $\rdy$ for each mass ratio sequence 
disappears for fairly stiff EOS.  We can see that the $\bJseq$ curves 
do not extend to the turning points for sequences with $N \sim 0.6$, 
$0.5$, $0.4$ and $0.3$ for $\ratmas = 1$, $0.5$, $0.2$ and $0.1$, 
respectively.  Similar to the Roche binary systems, we can roughly classify 
those sequences for the IRR binary systems in Table \ref{model-irr.tab} by 
considering the critical radii.  We will use three capital letters H, R and M 
as before.  For the IRR binary systems, the terminal configuration of the 
R-type sequences always has a cusp.  As mentioned before, the hydrodynamical 
instability would probably occur for the H-type sequences and Roche 
overflow would be expected for the R-type sequences.  From our results, the 
possibility of Roche overflow survives even for realistic ranges of
$\ratmas$ and $N$ if the GR effect is neglected.  
In Table \ref{irr-crit.tab}, we tabulate several physical quantities 
at the critical radii, $\rdy$ and $r_R$.  
In Figure \ref{irr-mass-radi.gra}, we plot these critical radii 
against the mass ratio.  In panels (a) and (b), we compare our results 
with those of the ellipsoidal approximation tabulated in LRS1 for polytropes 
with $N = 0$ and $1.0$.  For $N=0$ polytropes, the results are the same as 
those of Ury\=u \& Eriguchi (1998c).  On the other hand, the difference 
amounts to $10$--$15 \%$ for polytropes with $N=1$.  
In Figure \ref{roc-mass-radi.gra}(c), we show our results for polytropes 
with $N = 0, 0.5$ and $1.0$.  The dynamical instability limit disappears for 
the models with larger polytropic indices and with larger mass ratios 
as mentioned before.  
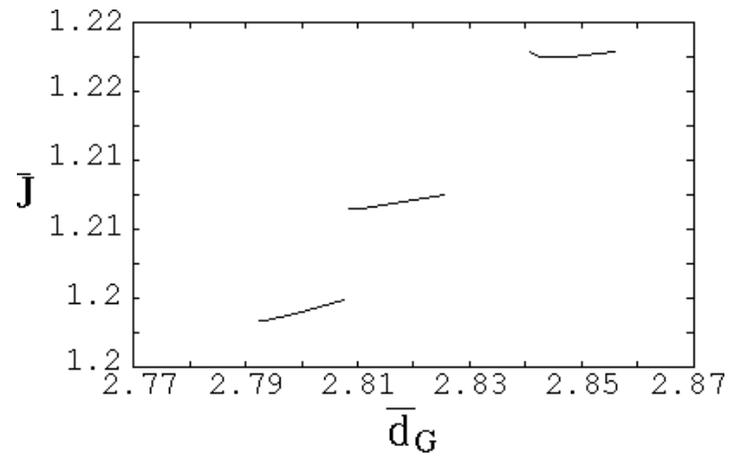
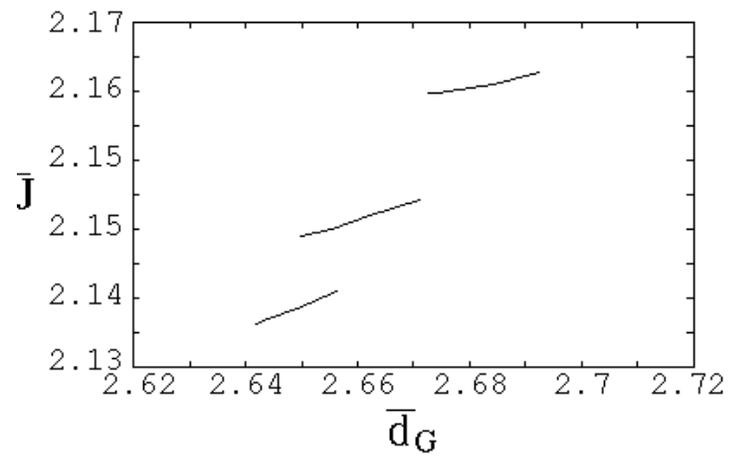
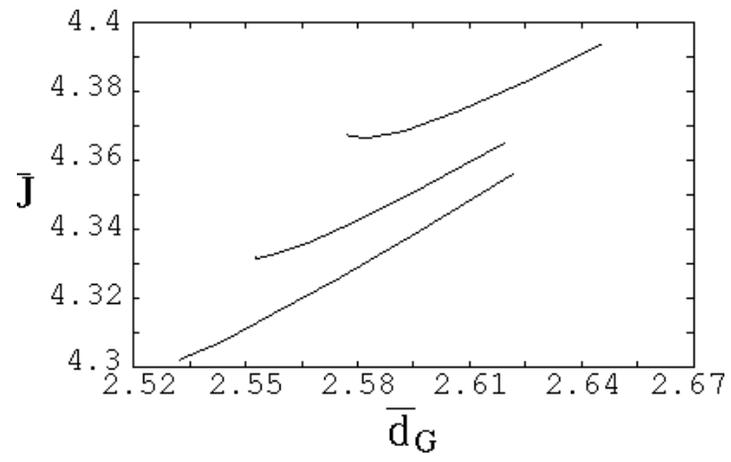
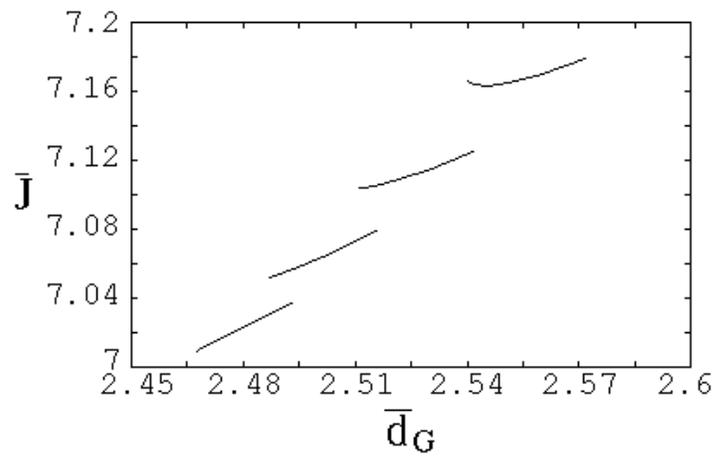
\begin{figure}
\vspace{20cm}
\caption{
Same as Fig.~1 but for the IRR sequences.
(a) $\ratmas = 1.0$ and $N=0.5$--$0.7$.  
(b) $\ratmas = 0.5$ and $N=0.5$--$0.7$.  
(c) $\ratmas = 0.2$ and $N=0.3$--$0.5$.  
(d) $\ratmas = 0.1$ and $N=0.2$--$0.5$.  }
\label{irr-d-J.gra}
\end{figure}
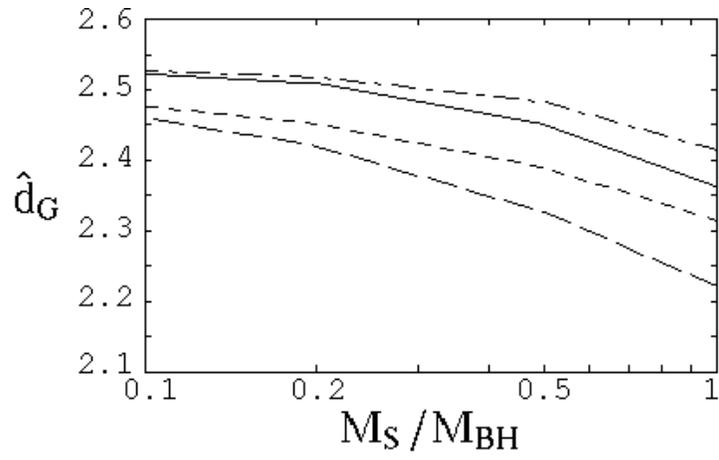
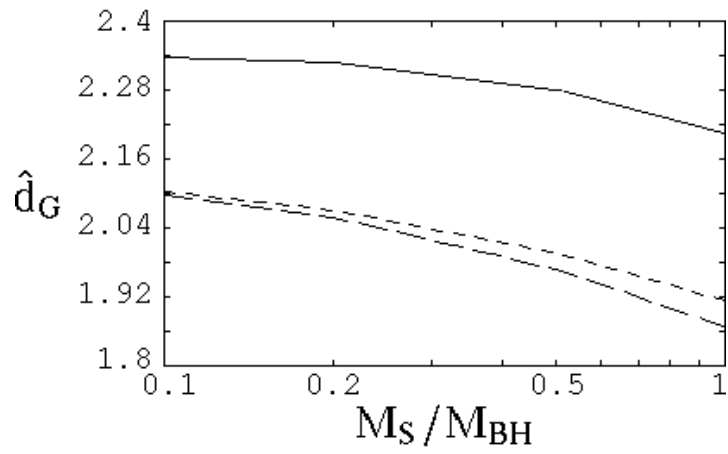
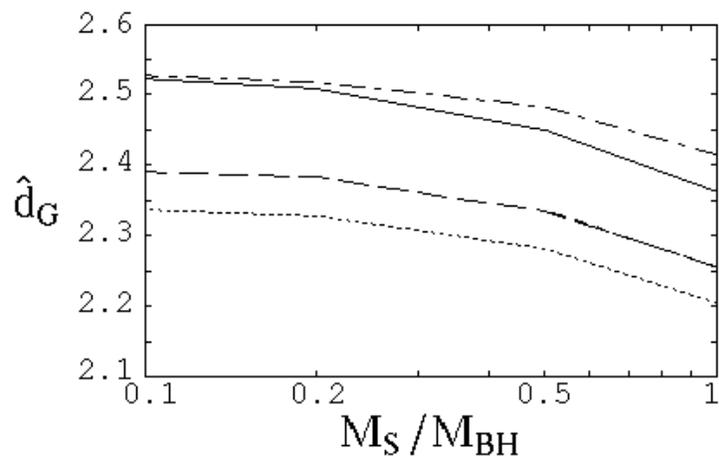
\begin{figure}
\vspace{15cm}
\caption{
Plots of critical radii $\rdy$ and $\rR$ for the IRR sequences as a function 
of the mass ratio $\ratmas$.  In panels (a) and (b), present results and those 
of LRS1 are shown for the polytropic indices $N=0$ and $1.0$,  respectively.  
Each curve corresponds to a different critical radius as follows:
present result for $\rdy$ (dash dotted line), present result for $r_R$ 
(solid line), result of LRS1 for $\rdy$ (short dashed line) and 
result of LRS1 for $r_R$ (long dashed line).  In the panel (c), 
present results for several polytropic indices $N = 0, 0.5$ and $1.0$ are 
drawn.  
Dash dotted line: $\rdy$ for $N=0$. 
Solid line: $r_R$ for $N=0$. 
Short dashed line: $\rdy$ for $N=0.5$.
Long dashed line: $r_R$ for $N=0.5$. 
Dotted line: $r_R$ for $N=1$. }
\label{irr-mass-radi.gra}
\end{figure}
\subsection{Configurations of the Roche type and the IRR type binary 
systems around the critical radii}

As shown in previous subsections, we have found that the locations of 
critical radii, in particular that of $r_R$, are quite different from those 
computed from the ellipsoidal approximation (LRS1).  Moreover, there is
a possibility that the fluid star reaches the Roche(--Riemann) limit without 
suffering from hydrodynamical instability because the critical radius such
as $\rse$ or $\rdy$ disappears on the stationary sequences for rather stiff 
EOS.  The reason for this difference of $r_R$ can be explained by considering 
the large deformation of the stellar envelope due to compressibility of the
stellar matter as follows (see also Paper I): the deformation of the envelope 
due to the tidal force becomes large, and, in particular, that around the 
inner parts near the BH becomes larger than that of other parts.  Because of 
this significant deformation of the envelope, a cusp is formed at the
inner edge and hence the sequence is terminated at a larger separation 
$\dg=r_R$ than that for the ellipsoidal approximation.  Therefore the 
difference between our results and those of LRS1 for the critical radii 
becomes significant.  
In Figures \ref{rocm10n05.con} and \ref{rrm10n05.con}, we show configurations 
of the fluid components for the Roche type and the IRR type binary systems, 
respectively.
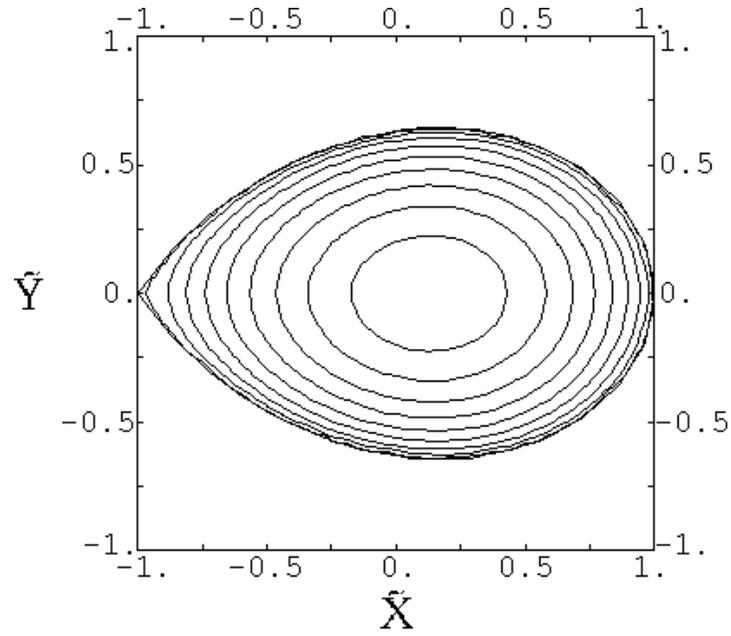
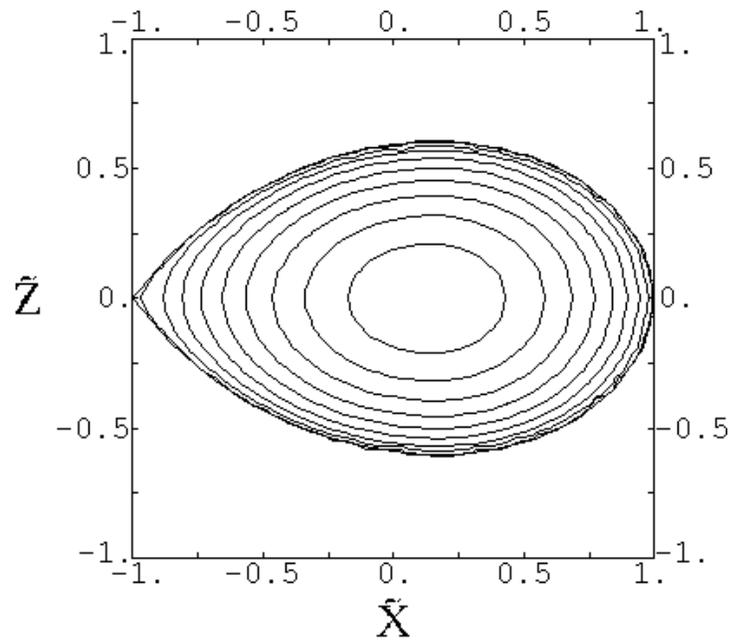
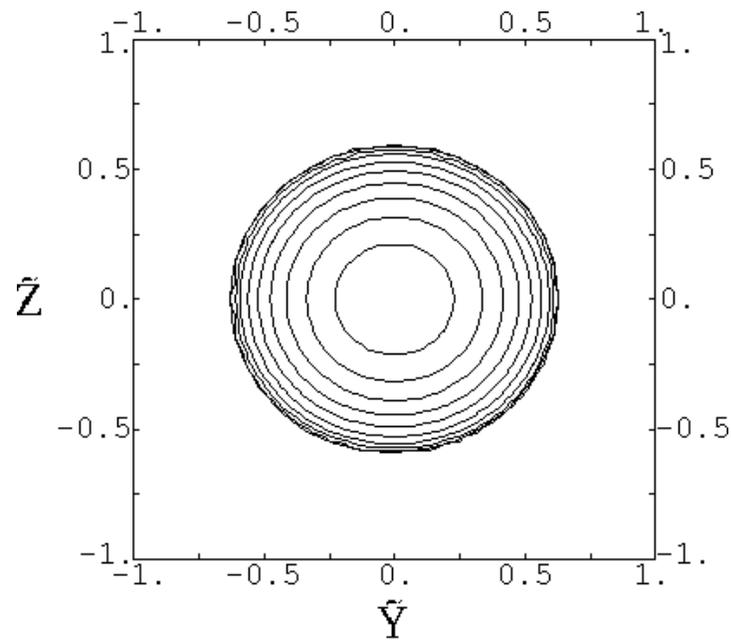
\begin{figure}
\vspace{19cm}
\caption{
Distributions of physical quantities for the Roche binary system 
with $\ratmas=0.1$ and $N=0.5$ polytrope at the Roche limit $\rR$.
$\tX \tY \tZ$--coordinates are the Cartesian coordinates 
where $\tX$ axis intersects the inner and the outer edges of the star 
and the rotational center.  $\tX$--$\tY$ plane is the 
equatorial plane.  The $\tZ$ axis is parallel to the rotational axis.  
The origin of $\tX \tY \tZ$--coordinates is at the 
geometrical center of the star and the coordinate is normalized by 
equation(\ref{rad}).  
(a) Contours of the density in the equatorial $\tX$--$\tY$ plane.
(b) Contours of the density in the meridional $\tX$--$\tZ$ plane.
(c) Contours of the density in the meridional $\tY$--$\tZ$ plane. 
The difference between two subsequent contours for each quantity is
$1/10$ of the difference between the maximum value and the minimum value.  }
\label{rocm10n05.con}
\end{figure}
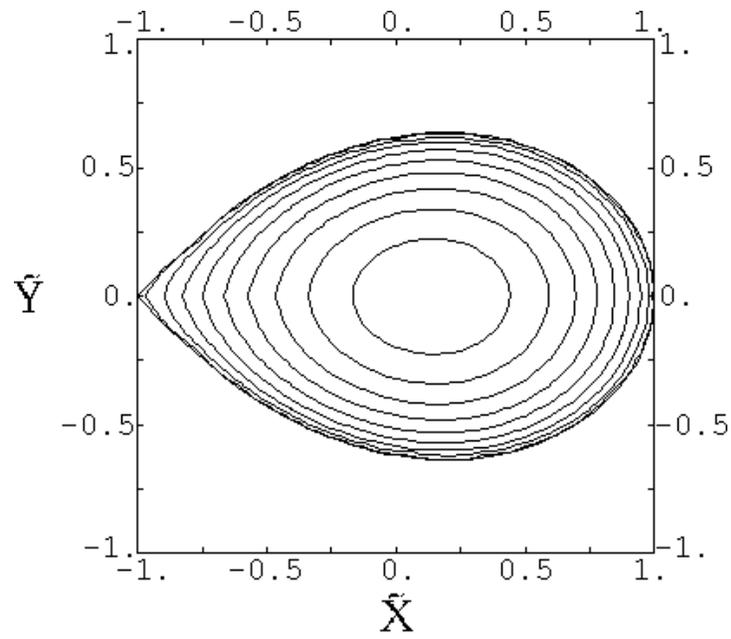
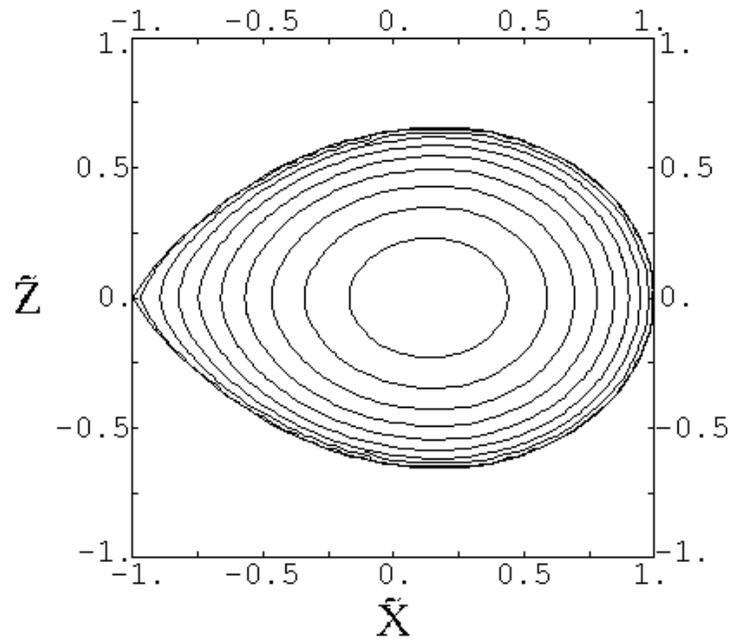
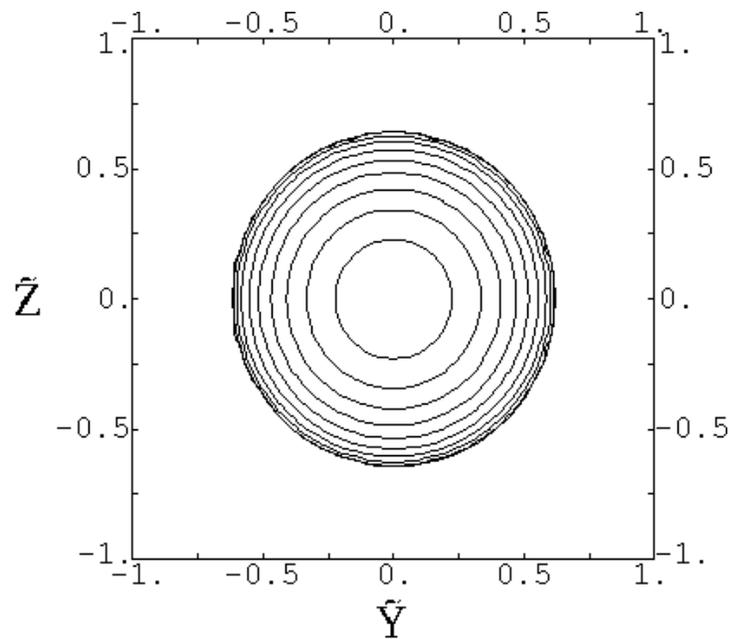
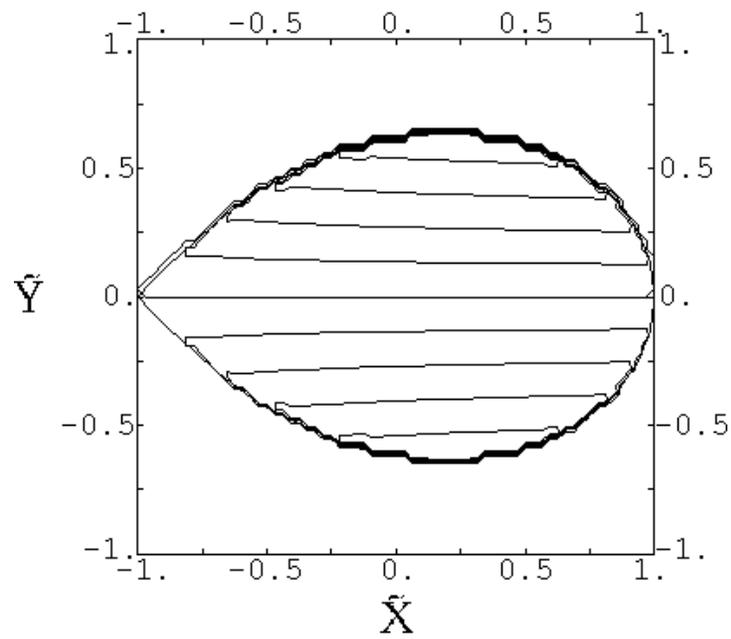
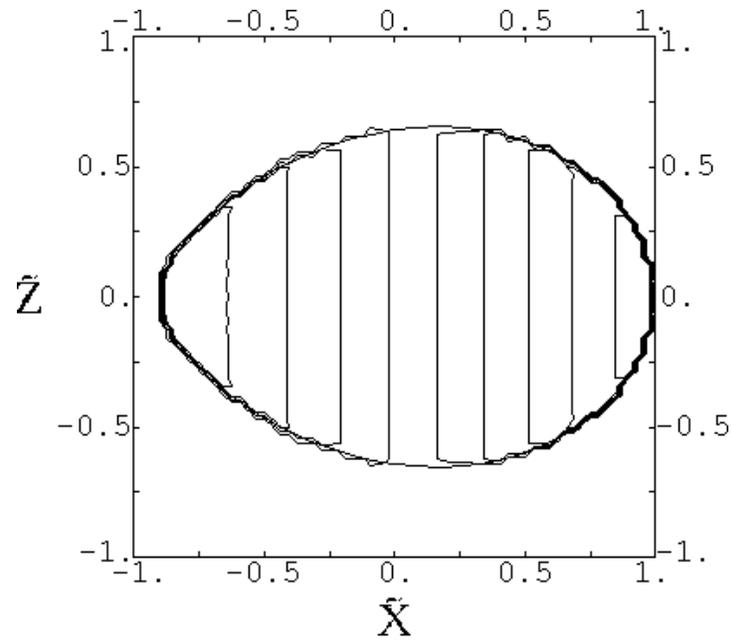
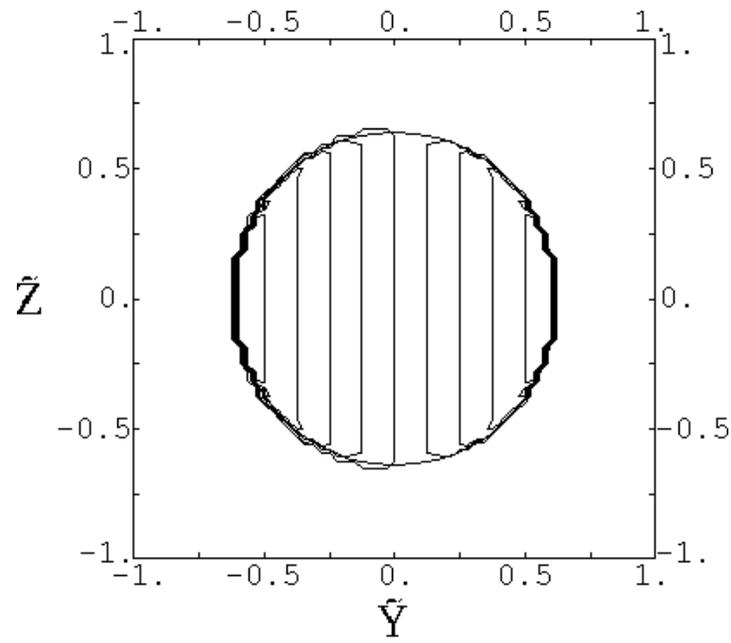
\begin{figure*}
\vspace{22cm}
\caption{
Distributions of physical quantities for the IRR binary system with 
$\ratmas=0.1$ and $N=0.5$ polytrope at the Roche-Riemann limit $\rR$.
(a) Contours of the density in the equatorial $\tX$--$\tY$ plane.
(b) Contours of the density in the meridional $\tX$--$\tZ$ plane.
(c) Contours of the density in the meridional $\tY$--$\tZ$ plane.
(d) Contours of the velocity potential in the equatorial $\tX$--$\tY$ plane.  
(e) Contours of the velocity potential in the plane with $\varphi=\pi/16$ 
and $\varphi=17\pi/16$.  
(f) Contours of the velocity potential in the meridional $\tX$--$\tZ$ plane.  
Conventions are the same as Figure \ref{rocm10n05.con}.  }
\label{rrm10n05.con}
\end{figure*}

From panels (a)-(c) of these figures, we can clearly see the difference of the 
shape in the two cases : the Roche type is oblate and the IRR type is prolate,
which is the same as the incompressible case.  Finally selected sequences of 
the Roche type and the IRR type binary systems near the critical points are 
tabulated in Tables \ref{roc-seq.tab} and \ref{irr-seq.tab}.  
\begin{table*}
\begin{minipage}{120mm}
\caption{Physical quantities for stationary sequences of the 
Roche binary systems.}
\label{roc-seq.tab}
\begin{tabular}{cccccccccccccc}
\hline
$\tilde{d}$ & $\bar{d}_G$ & $\bar{\Omega}$ & $\bar{J}$ &
$\bar{E}$ & $T/|W|$ & $\VC$ & $\bar{R}$ \\
\\[-2truemm]
\multicolumn{8}{c}{$\ratmas=1$ and $N = 0.5$} \\
\\[-2truemm]
2.4 & 2.942 & 3.286(-1) & 1.337 & -7.056(-1) & 1.893(-1) & 3.163(-2) & 1.020 \\
2.3 & 2.894 & 3.376(-1) & 1.335 & -7.063(-1) & 1.932(-1) & 3.130(-2) & 1.021 \\
2.2 & 2.856 & 3.453(-1) & 1.334 & -7.065(-1) & 1.968(-1) & 3.104(-2) & 1.023 \\
2.1 & 2.830 & 3.510(-1) & 1.335 & -7.063(-1) & 1.996(-1) & 3.092(-2) & 1.025 \\
2.05& 2.823 & 3.527(-1) & 1.335 & -7.060(-1) & 2.006(-1) & 3.096(-2) & 1.025 \\
2.0 & 2.820 & 3.534(-1) & 1.336 & -7.058(-1) & 2.011(-1) & 3.109(-2) & 1.026 \\
1.97& 2.821 & 3.532(-1) & 1.336 & -7.058(-1) & 2.010(-1) & 3.122(-2) & 1.026 \\
\\[-2truemm]
\multicolumn{8}{c}{$\ratmas=1$ and $N = 1.0$} \\
\\[-2truemm]
2.6 & 3.004 & 3.156(-1) & 1.312 & -6.539(-1) & 1.672(-1) & 3.333(-3) & 1.019 \\
2.5 & 2.942 & 3.260(-1) & 1.304 & -6.562(-1) & 1.708(-1) & 3.104(-3) & 1.022 \\
2.4 & 2.887 & 3.357(-1) & 1.298 & -6.581(-1) & 1.741(-1) & 3.335(-3) & 1.024 \\
2.3 & 2.840 & 3.445(-1) & 1.292 & -6.598(-1) & 1.771(-1) & 3.323(-3) & 1.026 \\
2.2 & 2.802 & 3.518(-1) & 1.288 & -6.609(-1) & 1.798(-1) & 3.415(-3) & 1.029 \\
2.1 & 2.780 & 3.562(-1) & 1.286 & -6.616(-1) & 1.814(-1) & 3.418(-3) & 1.030 \\
\\[-2truemm]
\multicolumn{8}{c}{$\ratmas=1$ and $N = 1.5$} \\
\\[-2truemm]
2.5 & 2.926 & 3.275(-1) & 1.278 & -5.895(-1) & 1.535(-1) & 1.358(-3) & 1.030 \\
2.4 & 2.871 & 3.371(-1) & 1.270 & -5.919(-1) & 1.564(-1) & 1.332(-3) & 1.033 \\
2.3 & 2.825 & 3.456(-1) & 1.263 & -5.940(-1) & 1.589(-1) & 1.414(-3) & 1.036 \\
2.2 & 2.790 & 3.522(-1) & 1.258 & -5.955(-1) & 1.608(-1) & 1.466(-3) & 1.039 \\
2.1 & 2.773 & 3.557(-1) & 1.255 & -5.963(-1) & 1.619(-1) & 1.536(-3) & 1.040 \\
\\[-2truemm]
\multicolumn{8}{c}{$\ratmas=0.1$ and $N = 0.5$} \\
\\[-2truemm]
4.0 & 2.495 & 3.097(-1) & 7.154 & -1.458 & 3.802(-1) & 1.311(-2) & 1.025 \\
3.9 & 2.480 & 3.128(-1) & 7.139 & -1.462 & 3.815(-1) & 1.310(-2) & 1.027 \\
3.8 & 2.467 & 3.155(-1) & 7.127 & -1.465 & 3.828(-1) & 1.313(-2) & 1.028 \\
3.7 & 2.457 & 3.176(-1) & 7.119 & -1.467 & 3.838(-1) & 1.320(-2) & 1.029 \\
3.6 & 2.451 & 3.189(-1) & 7.115 & -1.468 & 3.846(-1) & 1.336(-2) & 1.030 \\
3.5 & 2.449 & 3.194(-1) & 7.115 & -1.468 & 3.850(-1) & 1.363(-2) & 1.031 \\
\\[-2truemm]
\multicolumn{8}{c}{$\ratmas=0.1$ and $N = 1.0$} \\
\\[-2truemm]
4.0 & 2.435 & 3.198(-1) & 7.018 & -1.435 & 3.680(-1) & 2.258(-3) & 1.033 \\
3.9 & 2.421 & 3.228(-1) & 7.001 & -1.440 & 3.690(-1) & 2.246(-3) & 1.034 \\
3.8 & 2.410 & 3.251(-1) & 6.988 & -1.444 & 3.697(-1) & 2.242(-3) & 1.036 \\
3.7 & 2.404 & 3.264(-1) & 6.980 & -1.446 & 3.702(-1) & 2.260(-3) & 1.037 \\
3.6 & 2.404 & 3.262(-1) & 6.981 & -1.446 & 3.701(-1) & 2.252(-3) & 1.037 \\
\\[-2truemm]
\multicolumn{8}{c}{$\ratmas=0.1$ and $N = 1.5$} \\
\\[-2truemm]
4.3 & 2.469 & 3.127(-1) & 7.029 & -1.358 & 3.502(-1) & 6.646(-4) & 1.035 \\
4.2 & 2.448 & 3.169(-1) & 7.001 & -1.365 & 3.515(-1) & 6.709(-4) & 1.037 \\
4.1 & 2.428 & 3.207(-1) & 6.975 & -1.372 & 3.526(-1) & 6.937(-4) & 1.039 \\
4.0 & 2.412 & 3.239(-1) & 6.954 & -1.378 & 3.535(-1) & 6.947(-4) & 1.041 \\
3.9 & 2.399 & 3.265(-1) & 6.937 & -1.383 & 3.543(-1) & 7.055(-4) & 1.043 \\
3.8 & 2.392 & 3.282(-1) & 6.926 & -1.386 & 3.547(-1) & 7.292(-4) & 1.044 \\
\hline
\end{tabular}
\end{minipage}
\end{table*}
\begin{table*}
\begin{minipage}{120mm}
\caption{Physical quantities for stationary sequences of the 
IRR binary systems.}
\label{irr-seq.tab}
\begin{tabular}{cccccccccccccc}
\hline
$\tilde{d}$ & $\bar{d}_G$ & $\bar{\Omega}$ & $\bar{J}$ &
$\bar{E}$ & $T/|W|$ & $\VC$ & $\bar{R}$ \\
\\[-2truemm]
\multicolumn{8}{c}{$\ratmas=1$ and $N = 0.5$} \\
\\[-2truemm]
2.2 & 2.856 & 3.449(-1) & 1.223 & -7.239(-1) & 1.790(-1) & 2.159(-2) & 1.009 \\
2.15& 2.848 & 3.467(-1) & 1.222 & -7.240(-1) & 1.797(-1) & 2.163(-2) & 1.009 \\
2.1 & 2.842 & 3.479(-1) & 1.223 & -7.240(-1) & 1.802(-1) & 2.171(-2) & 1.010 \\
2.05& 2.841 & 3.484(-1) & 1.223 & -7.239(-1) & 1.805(-1) & 2.184(-2) & 1.010 \\
\\[-2truemm]
\multicolumn{8}{c}{$\ratmas=1$ and $N = 1.0$} \\
\\[-2truemm]
2.3 & 2.804 & 3.507(-1) & 1.194 & -6.765(-1) & 1.635(-1) & 3.177(-3) & 1.004 \\
2.25& 2.792 & 3.532(-1) & 1.191 & -6.772(-1) & 1.641(-1) & 3.159(-3) & 1.005 \\
2.2 & 2.783 & 3.550(-1) & 1.190 & -6.776(-1) & 1.645(-1) & 3.141(-3) & 1.005 \\
2.15& 2.778 & 3.560(-1) & 1.189 & -6.780(-1) & 1.648(-1) & 3.108(-3) & 1.005 \\
\\[-2truemm]
\multicolumn{8}{c}{$\ratmas=0.5$ and $N = 0.5$} \\
\\[-2truemm]
2.6 & 2.692 & 3.252(-1) & 2.164 & -8.419(-1) & 2.418(-1) & 1.712(-2) & 1.010 \\
2.55& 2.684 & 3.269(-1) & 2.163 & -8.424(-1) & 2.426(-1) & 1.715(-2) & 1.011 \\
2.5 & 2.678 & 3.282(-1) & 2.162 & -8.425(-1) & 2.433(-1) & 1.721(-2) & 1.011 \\
2.45& 2.674 & 3.291(-1) & 2.162 & -8.426(-1) & 2.437(-1) & 1.729(-2) & 1.011 \\
2.4 & 2.673 & 3.295(-1) & 2.162 & -8.426(-1) & 2.440(-1) & 1.741(-2) & 1.012 \\
\\[-2truemm]
\multicolumn{8}{c}{$\ratmas=0.5$ and $N = 1.0$} \\
\\[-2truemm]
2.6 & 2.620 & 3.361(-1) & 2.112 & -8.004(-1) & 2.263(-1) & 1.855(-3) & 1.007 \\
2.55& 2.614 & 3.372(-1) & 2.110 & -8.010(-1) & 2.266(-1) & 1.829(-3) & 1.007 \\
2.5 & 2.612 & 3.378(-1) & 2.109 & -8.012(-1) & 2.267(-1) & 1.774(-3) & 1.008 \\
\\[-2truemm]
\multicolumn{8}{c}{$\ratmas=0.2$ and $N = 0.5$} \\
\\[-2truemm]
3.4 & 2.575 & 3.096(-1) & 4.326 & -1.115 & 3.204(-1) & 1.074(-2) & 1.011 \\
3.3 & 2.558 & 3.131(-1) & 4.315 & -1.117 & 3.219(-1) & 1.068(-2) & 1.012 \\
3.2 & 2.544 & 3.159(-1) & 4.308 & -1.119 & 3.231(-1) & 1.064(-2) & 1.013 \\
3.1 & 2.535 & 3.177(-1) & 4.303 & -1.121 & 3.240(-1) & 1.065(-2) & 1.014 \\
3.0 & 2.532 & 3.184(-1) & 4.302 & -1.121 & 3.243(-1) & 1.072(-2) & 1.014 \\
\\[-2truemm]
\multicolumn{8}{c}{$\ratmas=0.2$ and $N = 1.0$} \\
\\[-2truemm]
3.6 & 2.543 & 3.137(-1) & 4.268 & -1.073 & 3.043(-1) & 1.739(-4) & 1.007 \\
3.5 & 2.518 & 3.184(-1) & 4.250 & -1.078 & 3.057(-1) & 9.681(-5) & 1.008 \\
3.4 & 2.498 & 3.224(-1) & 4.234 & -1.083 & 3.069(-1) & 2.461(-6) & 1.009 \\
3.3 & 2.483 & 3.254(-1) & 4.222 & -1.086 & 3.078(-1) & 1.047(-4) & 1.010 \\
3.2 & 2.474 & 3.273(-1) & 4.214 & -1.088 & 3.084(-1) & 2.389(-4) & 1.010 \\
\\[-2truemm]
\multicolumn{8}{c}{$\ratmas=0.1$ and $N = 0.5$} \\
\\[-2truemm]
4.0 & 2.493 & 3.106(-1) & 7.037 & -1.476 & 3.727(-1) & 6.340(-3) & 1.013 \\
3.9 & 2.482 & 3.129(-1) & 7.025 & -1.480 & 3.735(-1) & 6.258(-3) & 1.014 \\
3.8 & 2.474 & 3.146(-1) & 7.016 & -1.482 & 3.742(-1) & 6.191(-3) & 1.015 \\
3.7 & 2.469 & 3.157(-1) & 7.011 & -1.483 & 3.746(-1) & 6.145(-3) & 1.015 \\
3.65& 2.467 & 3.160(-1) & 7.009 & -1.484 & 3.747(-1) & 6.135(-3) & 1.015 \\
\\[-2truemm]
\multicolumn{8}{c}{$\ratmas=0.1$ and $N = 1.0$} \\
\\[-2truemm]
4.3 & 2.463 & 3.149(-1) & 6.956 & -1.441 & 3.589(-1) & 9.621(-4) & 1.009 \\
4.2 & 2.445 & 3.183(-1) & 6.933 & -1.448 & 3.598(-1) & 1.042(-3) & 1.010 \\
4.1 & 2.431 & 3.212(-1) & 6.913 & -1.453 & 3.605(-1) & 1.167(-3) & 1.011 \\
4.0 & 2.420 & 3.235(-1) & 6.898 & -1.457 & 3.611(-1) & 1.303(-3) & 1.011 \\
3.9 & 2.412 & 3.250(-1) & 6.887 & -1.460 & 3.614(-1) & 1.489(-3) & 1.012 \\
\hline
\end{tabular}
\end{minipage}
\end{table*}

\section{Discussion and conclusions}

\subsection{General relativistic effect and the possibility of mass transfer} 

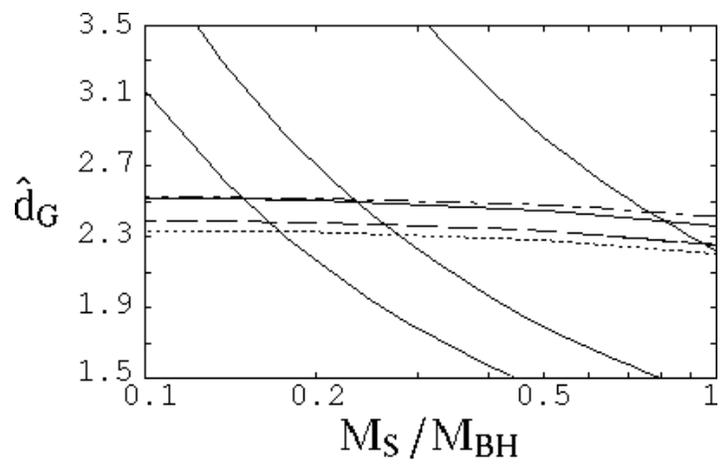
\begin{figure}
\vspace{5cm}
\caption{
Plots of the critical radii $\rdy$ and $\rR$ for the IRR sequences as a 
function of the mass ratio $\ratmas$.  Present results for polytropic 
indices $N = 0, 0.5$ and $1.0$ are the same as those drawn in Figure 
\ref{irr-mass-radi.gra}(c).  Three parallel curves drawn from upper 
left to lower right are those of $\hat{r}_\gr$ which correspond to
$G M_S / c^2 R_N=1/5$ (upper curve), $1/8$ and $1/10$ (lower curve). }
\label{irr-crit-rg.gra}
\end{figure}

So far, we have neglected the GR effect in order to focus on the 
hydrodynamical effect caused by the Newtonian tidal gravitational field.  For 
the realistic BH--star binary systems, the GR effect plays an essential role 
for the stability of the binary system when the star is sufficiently compact.  
Kidder, Will \& Wiseman (1992) developed a method to compute the radius of 
ISCO, $r_{GR}$, by taking the GR effect into consideration to some extent.  
They treated a system consisting of two point masses including the 
post-Newtonian effect up to $(v/c)^4$ order and also incorporated 
the effect of the Schwarzschild geometry exactly.  Lai, Rasio \& Shapiro 
proposed a fitting formula for their results in LRS2 as follows: 
\begin{equation}
r_\gr \,\simeq\, 
{G \over c^2}\,\left[\,6(M_S\,+\,M_\bh) \,+\, 
{4 \,M_S\,M_\bh \over M_S\,+\,M_\bh}\, \right] \, .
%%%\,=\, {G\,M_S \over c^2}\,\left[\,\,\right] 
\end{equation}
We use this formula to estimate roughly the GR effect.  In Figure 
\ref{irr-crit-rg.gra}, we draw the critical radii for the IRR sequence and 
the non-dimensional radius $\hat{r}_\gr$ which is defined as follows:  
\[
\hat{r}_\gr \,\simeq\, {G\,M_S\over c^2\,R_N}\,
{1 \over (1\,+\,\ratmasinv)^{1/3}}\, 
\]
\begin{equation}
\qquad \times \left[\,6\,\left(\,1\,+\,{1 \over \ratmas}\,\right)\,+\,
{4 \over 1\,+\,\ratmas}\, \right] \, ,
\end{equation}
where the normalization is the same as equation (\ref{idnorm}).
The factor $G M_S/c^2 R_N$ expresses the compactness of the neutron 
star.  We have chosen $G M_S / c^2 R_N=1/5$, $1/8$ and $1/10$.  
Values of $1/5$ and $1/10$ almost correspond to those of typical neutron 
stars with, say,  $M_S=1.35\, M_\odot$, $R_N=10\,{\rm km}$, 
and $M_S=0.68\, M_\odot$, $R_N=10\,{\rm km}$, respectively.  

In this figure, three parallel curves drawn from upper left to lower 
right are those of $\hat{r}_\gr$ and they correspond to $G M_S/c^2 R_N
=1/5$, $1/8$ and $1/10$ from the top to the bottom curves, respectively.  
We can see that $r_\gr$, which is the instability limit due to the GR effect, 
can be comparable with the Roche-Riemann radius $\rR$ for a binary system 
with a less compact neutron star with $G M_S/c^2 R_N\sim1/8$ and moderate 
mass ratios.  Therefore the configuration at the Roche-Riemann limit $r_R$ 
could be stable even if the GR effect is included and hence the mass overflow 
instead of unstable plunge is expected to occur.  The same mechanism can be 
applied for the Roche type binary systems, because, as we can see from 
Tables \ref{roc-crit.tab} and \ref{irr-crit.tab}, the values of $r_R$ are 
almost the same both for the Roche binary systems and for the IRR 
binary systems.  

Taniguchi \& Nakamura (1996) have investigated stability of the Roche type 
binary systems by using the improved pseudo-Newtonian potential and the 
ellipsoidal approximation.  Because of the ellipsoidal approximation, the 
hydrodynamical instability limit always appears on the solution sequence 
of their result as LRS1, which is in contrast to our present result.  
Recently, Klu\'zniak \& Lee (1997) performed numerical simulations for binary
systems of a point source and a synchronous rotating star by using the SPH 
method.  They reported models with mass ratios $\ratmas = 1$ and $0.31$ for 
polytropes with $N=0.5$.  In their paper, they have shown that not the 
unstable plunge but the mass transfer occurs for $\ratmas < 0.8$.  
The resulting mass transfer from NS to BH and matter distribution around 
the BH vary significantly.  They have concluded that the system is a promising 
source of GRB.  Our computation qualitatively agrees with their result.  For 
the Roche binary with $N=0.5$, our result shows that $\rse$ disappears around 
$\ratmas \la 0.2$.  Although this value is different from the result of 
simulations, values of radii $\rse$ and $\rR$, and hence $\rdy$ are 
very close as seen in Figure \ref{roc-mass-radi.gra}(d).  Therefore 
the possibility of mass overflow is also supported in this respect.  

If such a mass overflow occurs, we need to consider the following problem.
Does this mass transfer continue stably or not?  Stable mass 
transfers lead to the survival of a less massive NS and the BH--NS binary 
systems 
with a much smaller mass ratio $\ratmas$ are formed after the mass transfer.  
This process depends on several factors such as the amount of mass escaped 
from the NS and that accreted onto the BH, and on how the structure of NS and 
its Roche lobe have changed as a result of overflow and so on.  Qualitative 
discussions (see Kochanek 1992 ; Bildsten \& Cutler 1992, and references 
therein) suggest 
that such stable mass transfer seems difficult.  Quantitative examination  
is possible by using our stationary method.  We will investigate this problem 
in the future.  

\subsection{GW signal at the Roche(--Riemann) limit}

Let us consider the case that the BH--NS binary evolves stably along the 
stationary sequence and reach the Roche(-Riemann) limit $\rR$ without 
suffering from hydrodynamical instability.  It is natural to expect 
that the GW signal emitted from such a binary system would change at 
a critical radius and from the frequency there we may be able to determine 
the critical radius, $r_R$ in our case.  

By comparing Tables \ref{roc-crit.tab} and \ref{irr-crit.tab}, the physical 
quantities around the critical radii are almost the same for the Roche and 
the IRR type binary systems.  In particular, the difference of $r_R$ 
between the Roche and the IRR types is less than $2, 1$ and $0.5 \%$ for 
polytropes with $N=0, 0.5$ and $1$, respectively.  This is irrespective of 
the mass ratio for the range $0.1 \la \ratmas \la 1$.  Therefore, we cannot 
decide whether the star is in a synchronously rotating state or in an 
irrotational state from the observation of the GW.  In other words, 
the GW signal contains information about EOS but not about the rotational 
state.  Therefore we can only determine $r_R$ and the stiffness of the NS.  
On the other hand, it seems difficult to extract some new information for 
the viscosity of neutron matter from the GW signal emitted at this stage.  
Note, however, the spin of NS may be accurately determined by using the 
higher order PN expansion for the inspiraling phase (see for e.g. 
Mino et. al. 1997, and references therein).  

\subsection{Conclusions}

In this paper we have systematically examined the sequences of a point 
source and a fluid star binary systems.  We have assumed the fluid star to be 
a polytrope and employed Newtonian gravity.  We have found that the critical 
radii, $\rdy$, $\rse$ and $r_R$ are quantitatively different from those 
computed by using the ellipsoidal approximation for the star.  We have 
clarified that $\rdy$ or $\rse$ disappears even on binary sequences with a 
moderate mass ratio and a stiff EOS and hence the existence of dynamically 
stable Roche(-Riemann) limit has been shown.  This suggests a possibility of 
mass overflow instead of an unstable plunge due to the hydrodynamical 
instability of binary systems at the final stage of the binary evolution due
to the GW emission.  

For further investigations, we should take into account effects of the black 
hole field, the radial velocity and so on.  Though it is very difficult to 
numerically construct the realistic BH, the pseudo-Newtonian potential can 
represent the marginal instability of the GR effect around the Schwarzschild 
BH field.  Taniguchi \& Nakamura (1996) have used an improved pseudo-Newtonian 
potential and the ellipsoidal approximation to determine the innermost stable 
circular orbit of BH-NS systems.  Such a pseudo-Newtonian potential will be 
included straightforwardly in our present scheme for a fully deformed 
configuration of stars.  

Newtonian approximation for the fluid star component is insufficient 
for NS models whose gravity is considered as strong as, 
more or less, the second post-Newtonian approximation of general 
relativity.  Recently several authors have solved the NS-NS 
configuration including the GR effect (Shibata 1997 ; Baumgarte et al. 1997).
The formulation for the irrotational star in GR gravity is 
also proposed (Bonazzola, Gourgoulhon \& Marck 1997 ; Asada 1998 ; 
Shibata 1998 ; Teukolsky 1998).  
By implementing these results, the GR effect could be included step by step, 
and finally it will be possible to construct the quasi-stationary 
evolution of the BH--NS systems.  

\section*{Acknowledgments}

We would like to thank Prof. J. C. Miller for carefully reading the 
manuscript and for helpful comments.  KU would also like to 
thank Prof. D. W. Sciama and Dr. A. Lanza for their warm hospitality at 
ICTP and SISSA.  Numerical computations were carried out by at the 
Astronomical Data Analysis Center of the National Astronomical 
Observatory, Japan.  
{}
\end{document}